\documentclass[fleqn,usenatbib]{mnras}

\usepackage{newtxtext,newtxmath}

\usepackage[T1]{fontenc}

\DeclareRobustCommand{\VAN}[3]{#2}
\let\VANthebibliography\thebibliography
\def\thebibliography{\DeclareRobustCommand{\VAN}[3]{##3}\VANthebibliography}

\usepackage{graphicx}	
\usepackage{amsmath}	

\usepackage[caption=false]{subfig}
\usepackage [english]{babel}

\def\mevs{\hbox{\rm\hskip.35em  MeV$^{-1}$ s}$^{-1}$}
\def\hzs{\hbox{\rm\hskip.35em  Hz s}$^{-1}$}
\def\cmss{\hbox{\rm\hskip.35em  cm$^{2}$ s}$^{-1}$}

\title[PSR J0537 and Vortex Creep Model]{Glitch analysis and braking index determination of the unique source PSR J0537$-$6910}

\author[Akbal et al.]{
Onur Akbal,$^{1}$\thanks{E-mail: akbalonur85@gmail.com (OA)}
Erbil G\"{u}gercino\u{g}lu,$^{2}$\thanks{E-mail: egugercinoglu@gmail.com (EG)}
M.~Ali Alpar$^{2}$\thanks{E-mail: ali.alpar@sabanciuniv.edu (MAA)}
\\
$^{1}$Bah\c{c}e\c{s}ehir College, \c{C}i\c{c}eklik\"{o}y, Bornova, 35040 Izmir, Turkey\\
$^{2}$Faculty of Engineering and Natural Sciences, Sabanc{\i} University, Orhanl{\i}, Tuzla, 34956 Istanbul, Turkey\\
}

\date{Accepted XXX. Received YYY; in original form ZZZ}

\pubyear{2021}

\begin{document}
\label{firstpage}
\pagerange{\pageref{firstpage}--\pageref{lastpage}}
\maketitle

\begin{abstract}
PSR J0537$-$6910 is the most active glitching pulsar with a glitch rate $\sim0.3$ \mbox{yr$^{-1}$}. We have reanalysed 45 glitches of PSR J0537$-$6910 published in the literature and have done post-glitch timing fits within the vortex creep model. Moment of inertia fractions of the superfluid regions participating in glitches are obtained for each event and the model predictions for the inter-glitch time are confronted with the observed time-scales. Similarities and differences with the glitching behaviours of the well studied Crab and Vela pulsars are highlighted. From superfluid recoupling time-scales we estimate an inner crust temperature of $T=0.9\times10^{8}$ K for PSR J0537$-$6910. It is found that PSR J0537$-$6910 glitches leave behind persistent shifts similar to those observed from the Crab pulsar. These persistent shifts are responsible for the long term increase of the spin-down rate of PSR J0537$-$6910 and the apparent negative trend in its braking. Glitch magnitudes and persistent shifts of PSR J0537$-$6910 are consistent with the scenario that this is a young pulsar in the process of developing new vortex traps. We determine a braking index $n=2.7(4)$ after glitch induced contributions to the rotational evolution have been removed.
\end{abstract}




\section{Introduction}
\label{sec:intro}
The X-ray pulsar PSR J0537--6910 in the Large Magellanic Cloud (LMC) is a unique source in terms of its rotational properties. Its spin evolution is interrupted by large glitch events, i.e. abrupt changes in the rotation and spin-down rates, followed by post-glitch relaxations. This is the most actively glitching pulsar, with the rate of $ \gtrsim 3.2 $ per year, and the magnitudes $\Delta\nu$ of its glitches are the largest observed among young pulsars \citep{middleditch06,antonopoulou18,ferdman18,ho20}. There is a vigorous correlation reported between the glitch magnitude and the time until the next glitch for this pulsar. The domination of large glitches in the rotational history provides a unique opportunity to study the contribution of the  interior dynamics of the neutron star. This remarkable regularity of the glitches, not observed to this extent in other pulsars, can be interpreted in terms of a combination of crust-quakes and superfluid dynamics \citep{baym71,ruderman76,alpar96}.

The large effect of the glitch activity of this pulsar on the spin evolution also makes the measurement of its braking index, $ n=\nu \ddot{\nu}/ \dot{\nu} $, quite ambiguous. \citet{middleditch06} and \citet{antonopoulou18} have found a \textit{negative} braking index from the overall increase tendency of the spin-down rate in the long run, while \citet{ferdman18} have inferred a short term braking index of $ n\simeq 7.4 $ from data between successive glitches.

Our purpose is twofold. Firstly, we make detailed fits to the glitches of PSR J0537--6910 to obtain superfluid characteristics of this pulsar and to constrain neutron star structural parameters. Secondly, we present a method to find the true braking index for frequently glitching pulsars and infer the underlying spin-down law of PSR J0537--6910.
   
We investigate the post-glitch timing data of PSR J0537--6910 in terms of the vortex creep model. In \S\ref{sec:observations} we overview the timing of PSR J0537--6910 and summarize the literature on its glitch properties and braking index measurements. In \S\ref{sec:modelfits} we present our fits to the post-glitch timing data after a summary of the vortex creep model. In \S\ref{sec:braking} we present the method for braking index determination. In \S\ref{sec:results} we use the results of the model fits to constrain neutron star internal structure and dynamics. And finally in \S\ref{sec:disandconc} we discuss implications of our results and compare the properties of PSR J0537--6910 with the Crab and Vela pulsars. 

\section{Literature overview for the timing of PSR J0537$-$6910}
\label{sec:observations}
PSR J0537--6910 resides in the 1-5 kyr old supernova remnant N157B \citep{chen06}, which is 49.6 kpc away from the Earth \citep{pietrzynski19}. It was discovered in the Large Magellanic
Cloud by the Rossi X-ray Timing Explorer (RXTE) \citep{marshall98}. PSR J0537--6910 is the fastest rotating young pulsar, with spin frequency $ \nu=62 $ Hz and has very high spin-down rate ($ \dot{\nu}=-1.992 \times 10^{-10}$ Hz s$^{-1}$), giving the largest spin-down luminosity with $ \dot{E}=4.9 \times 10^{38} $ erg s$^{-1}$. It has a characteristic age $ \tau_c=4.9 $ kyr and inferred surface dipole magnetic field $ B_s=9.25 \times 10^{11} $ G. 

PSR J0537--6910 immediately exhibited the evidence of large glitch activity in its first timing observations \citep{marshall98,marshall04}. Using X-ray data obtained from RXTE, \citet{middleditch06} reported the follow up timing observations over 7.5 years, including 24 glitches. In this comprehensive analysis they determined some conspicuous timing features of this source: (i) This is the pulsar of the largest glitch rate, 3 yr$^{-1} $. (ii) There is a strong correlation between the glitch sizes and the time until the following glitch with a correlation coefficient of $ \sim 0.94 $, not observed in other pulsars. (iii) The long-term tendency of spin-down rate, $\left|\dot{\nu}\right|$, to  increase over time gives the long-term \textit{negative} braking index, $ n=-1.5 $. (iv) There is a correlation between the time interval preceding a glitch and the change in the spin-down rate, $ \Delta \dot{\nu} $. (v) Microglitches show up preceding the large glitches. 

With a longer (13 year of) data string from RXTE, \citet{antonopoulou18} and \citet{ferdman18} recently presented timing analyses in which they identified 45 and 42 glitches respectively. Their combined results are as follows:

\begin{itemize}
	\item There is no explicit correlation between sizes of the change in frequency, $ \Delta \nu $, and change in spin-down rate, $ \Delta \dot{\nu} $.
	\item Glitch sizes, $ \Delta \nu $, spanning about two orders of magnitude, seem to be normally distributed, while in other pulsars the distribution is well-described by a power law \citep{melatos08,espinoza14}.
	\item There is a strong linear correlation between the glitch size and waiting time until the subsequent glitch, with a slope of $ \sim 0.2 $ day$ ^{-1} $, while the glitch size is not correlated with the time preceding the glitch.
	\item The changes in spin-down rate, $ \Delta \dot{\nu} $, do not appear to be correlated with the preceding and subsequent inter-glitch times, $ t_{g}$.
	\item The activity parameter describing the cumulative frequency change over the observation time-span, 
	 \begin{equation}
A_{\rm g}=\frac{1}{T_{\rm obs}}\sum_{i=1}^{N_{\rm g}} \left(\frac{\Delta\nu}{\nu}\right)_{i},
\label{gactivity}
\end{equation}
of this pulsar is found to be $A_{\rm g}=9 \times 10^{-7}$ yr$^{-1}$  $T_{\rm obs}$ being the total observation time containing $N_{\rm g}$ glitches. PSR J0537--6910 has the second highest glitch activity parameter after PSR J1023--5746 which has $A_{\rm g}=14.5\times10^{-7}$ yr$^{-1}$ \citep{erbil21}.
	\item The inter-glitch evolution of the spin-down rate is roughly described by a linear increase with the slope of $ \ddot{\nu} \sim 10^{-20}$ Hz s$ ^{-1} $ giving anomalous interglitch braking index $ n_{ig} \sim 20 $ on the average \citep{antonopoulou18,ferdman18}. \citet{ferdman18} particularly analyzed the post-glitch $ \dot{\nu} $ evolution, applying Markov Chain Monte Carlo (MCMC) fits to all interglitch data segments. Their phenomenological model contains linear (with the slope of $ m=7.6 \times 10^{-16} $ s$ ^{-2} $), and exponential terms (with a decaying time $ \tau \approx 27 $ days and amplitude $ A \approx 7.6 \times 10^-14 $ s$ ^{-2} $). As we show in this work such a procedure gives misleading conclusions on the braking index value of the pulsar; see \S\ref{sec:braking} below for details.
	\item The high glitch activity of this source makes its long term braking index measurement quite ambiguous. In contrast to the interglitch behaviour, there is an overall trend of increase in the spin-down rate in the long run, implying a negative long-term braking index, $ n \simeq -1.2 $ \citep{antonopoulou18}. Claiming that this uncommon measurement is caused by the artificial effect of the high glitch activity, \citet{ferdman18} inferred the overall braking index as $ n \simeq 7 $ from the slope of the linear evolution with time of the spin-down rate in their inter-glitch MCMC fit. This value was interpreted as the spinning down of the neutron star owing to gravitational waves from an unstable r-mode \citep{andersson18}.
	\item The emission properties, i.e. flux variations, pulse profiles, burst activity, have not displayed changes associated with the glitches \citep{ferdman18}.
\end{itemize}
After the decommissioning of RXTE in 2012 PSR J0537--6910 was observed with the Neutron Star Interior Composition Explorer (NICER) for the period 2017 August 17 and 2020 October 29. During this timespan 11 new glitches were discovered \citep{ho20,abbott20}. These new glitches have the same properties as the glitches observed with RXTE; in particular the glitch magnitudes are correlated with the time to the next glitch.  

\section{Post-glitch Timing and Model Fits}
\label{sec:modelfits}

\subsection{Overview of the Vortex Creep Model}

The exchange of angular momentum between the superfluid components of 
a neutron star and its observed outer crust is governed by the distribution and 
motion of quantized vortices in the superfluid. The superfluid core is 
tightly coupled to the normal matter due to electron scattering off 
magnetized vortices on very short time-scales \citep{ALS84}. 
Hence the superfluid component in the core which comprises the bulk of the star 
effectively rotates together with the observed outer crust.

However, vortex movement is obstructed in other superfluid parts of the 
star by pinning to inhomogeneties such as the nuclei in the inner 
crust superfluid \citep{alpar77} or toroidal flux lines in the outer 
core superconductor \citep{erbil14}. 
Sudden unpinning events are the main mechanism responsible for 
the angular momentum transfer to the crust observed as glitches 
\citep{anderson75}. In between the glitches pinned superfluid 
components manage to spin down continuously by thermally activated 
motion of vortex lines over pinning energy barriers, i.e. the process of 
vortex creep \citep{alpar84,alpar89}. The interglitch timing behaviour 
of a pulsar is explained by the response of the creep process to glitch 
induced resets of the superfluid and normal crust rotation rates. 
The creep response is intrinsically highly nonlinear as it depends 
exponentially, through Boltzmann factors, on the lag between the 
rotation rates of the superfluid and normal matter components of the 
neutron star. The linear response limit of the creep  process prevails in 
some parts of the neutron star,  leading to early exponential relaxation of part of the glitch \citep{alpar89,erbil17}. Full nonlinear response manifests itself as power law 
post-glitch and interglitch behaviour. The 
basic nonlinear response is the stopping of the creep 
process completely for a while until it resumes after a certain waiting 
time. The vortex creep model, with linear and nonlinear 
regimes of response to a glitch, provides the physical context 
for understanding both short term exponential relaxation and long-term 
power-law inter-glitch timing. Here we summarize the relevant concepts 
of the model and refer to earlier work for details \citep{alpar84,alpar89,alpar96,akbal15,erbil19,erbil20}.

A non-linear creep region $k$, with moment of inertia 
$I_{\rm s, k} $ contributes to the postglitch response of the 
observed spin-down rate with a term of the Fermi function form \citep{alpar84}
\begin{equation}
\Delta\dot\nu (t)=\frac{I_{\rm s, k}}{I}\dot\nu_{0}\left[1-\frac{1}{1+({\rm e}^{t_{\rm s,k}/\tau_{\rm nl}}-1){\rm e}^{-t/\tau_{\rm nl}}}\right],
\label{creepsingle}
\end{equation}
where $I$ is the total moment of inertia of the star and $\dot\nu_{0}$ 
is the fiducial spindown rate right before the glitch. Vortex creep stops temporarily in response to 
the glitch induced changes 
in superfluid and crust rotation rates. This particular creep region is now 
not contributing to angular momentum transfer to the crust whose 
spin-down rate therefore increases by $|\Delta\dot\nu (t)|$. 
Creep resumes and $\Delta\dot\nu (t)$ relaxes quickly to zero 
after a waiting time 
$t_{\rm s,k} \cong \delta\Omega_{\rm s,k}/\left|\dot\Omega\right|$ , 
within a time interval $\tau_{\rm nl}$,  
the nonlinear creep relaxation time defined as
\begin{equation}
\tau_{\rm nl}\equiv \frac{kT}{E_{\rm p}}\frac{\omega_{\rm cr}}{\vert\dot{\Omega}\vert},
\end{equation}
where $\omega_{\rm cr}$ is the critical lag for unpinning and $E_{\rm p}$ is the pinning energy for vortex line-lattice nuclei interaction, $T$ is the inner crust temperature and $\left|\dot\Omega\right|$ is the absolute magnitude of the spin-down rate.

This most extreme nonlinear signature of stop-hold-and-restart, 
the basic ingredient of nonlinear response, is rarely discernible in the observed 
spin-down rate, if $I_{\rm s, k}/I$ is large enough for a creep region 
effected by a single constant value of $\Omega_{\rm s,k}$. 
This signature was observed first in the Vela pulsar by \citet{buchner08} and then also detected in PSRs J1023$-$5746, J2111+4606, and J2229+6114 \citep{erbil21}. It is also discerned after glitches 1, 16, 24 and 45 of PSR J0537--6910 (see supplementary material), as we find below. 

More commonly, the glitch 
effect on the superfluid rotation rate $\Omega_{\rm s}$ should vary continuously over nonlinear 
creep regions, leading to a sequence of small step recoveries with 
consecutive recovery times \citep{alpar84,erbil20}: 
\begin{equation}
\noindent
\frac{\Delta\dot{\nu}(t)}{\dot\nu}=\frac{I_{A}}{I}
\left\lbrace1-\frac{1-(\tau_{\rm nl}/t_{0})\ln\left[1+(e^{t_{0}/\tau_{\rm nl}}-1)e^{-\frac{t}{\tau_{\rm nl}}}\right]}{1-e^{-\frac{t}{\tau_{\rm nl}}}}\right\rbrace.
\label{creepfull}
\end{equation}

In Eq.(\ref{creepfull}) $I_{\rm A}$ denotes the total moment of inertia of the 
non-linear creep regions contributing, $t$ is the time measured from the glitch date and $ t_0 $ is the maximum 
waiting time, which is set by the maximum, $\delta\Omega_s$, of the decrease  
in the local superfluid rotation rate due to the sudden outward motion of the 
vortices at the time of the glitch,
\begin{equation}
t_0 = \frac {\delta\Omega_{\rm s}}{\left|\dot\Omega\right|}.
\label{waitingtime}
\end{equation}
The contribution of the offset in nonlinear creep to the glitch associated change in the 
spin-down rate is  
\begin{equation}
\frac{\Delta \dot \nu(0)}{\dot \nu}=\frac{I_{\rm A}}{I}.
\label{glitchdotnu0}
\end{equation} 

The cumulative result, Eq.(\ref{creepfull}), reduces to the 
commonly observed interglitch recovery of the spin-down rate with a 
constant second derivative of the rotation rate 
\begin{equation}
\frac{\Delta \dot \nu(t)}{\dot \nu}=\frac{I_{\rm A}}{I}\left(1-\frac{t}{t_{0}}\right),
\label{glitchdotnu}
\end{equation}
when $ t_0 \gg \tau_{\rm nl} $.
The glitch induced decrease $\delta\Omega_{\rm s}$ in the angular velocity of 
the superfluid is related to the number $N_{\rm vortex}$ 
of unpinned vortices moving outward through the superfluid at the glitch,
\begin{equation}
N_{\rm vortex}=2\pi R^{2}\frac{\delta\Omega_{\rm s}}{\kappa},
\label{Nvortex}
\end{equation}
where $R$, the distance from the rotation axis, is of the order of the 
neutron star radius for the crustal superfluid region.
     
The interglitch behaviour given in Eq.(\ref{glitchdotnu}) is observed in Vela glitches \citep{akbal17}, in Vela-like pulsars \citep{espinoza17,erbil21} 
and other giant glitches in older pulsars \citep{yu13}. The full expression given in Eq.(\ref{creepfull}) for the cumulative response was needed for fitting the postglitch response following glitches 1, 16, 24 and 45, for which enough data was available at the initial and final stages of the relaxation. The constant 
interglitch frequency second derivative ($\ddot{\nu}$) given in 
Eq.(\ref{glitchdotnu}) is related to the parameters of the glitch
\begin{equation}
\ddot \nu_{\rm ig}=\frac{I_{\rm A}}{I}\frac{|\dot\nu|}{t_{0}},
\label{ddotnuig}
\end{equation}
and is much larger than the expected long-term second derivative associated with the dipole external torque on the pulsar. 
The `anomalous' braking indices observed from many pulsars 
\citep{johnston99,dang20} can be understood as due to 
such interglitch behaviour due to nonlinear creep response 
\citep{alpar06}. 

The vortices  unpinned at a glitch also move 
through some regions of the superfluid, of total moment of inertia 
$I_{\rm B}$, which ordinarily do not contain vortices and therefore do 
not contribute to creep and to the spin-down rate.
Equations (\ref{glitchdotnu0}) and (\ref{ddotnuig}) are supplemented by 
angular momentum conservation relating the observed step increase 
in the pulsar rotation frequency to the decrease in superfluid rotation rate 
due to the sudden motion of vortices unpinned in the glitch
\begin{equation}
I_{\rm c}\Delta \Omega_{\rm c} = (I_{\rm A}/2 + I_{\rm B}) \delta \Omega_{\rm s},
\label{glitchomega}
\end{equation}
where the prefactor 1/2 accounts for a linear decrease of the 
superfluid angular velocity within non-linear creep regions with 
moment of inertia $I_{\rm A}$. The effective crust moment of inertia $I_{\rm c}$ 
includes the core superfluid which couples to the normal matter on timescales short 
compared to the resolution of the glitch rise; hence $I_{\rm c} \cong I$, 
the total moment of inertia of the star. The first indication of a limit < 12 s on the glitch rise 
\citep{ashton19} was interpreted as a limit on the crust core coupling time \citep{erbil20}.

The waiting time $t_0$ for the recovery of the nonlinear creep response can be obtained from the glitch 
associated step in the spin-down rate and the observed interglitch $\ddot{\nu}$ using 
Eqs.(\ref{glitchdotnu0}) and (\ref{ddotnuig}): 
\begin{equation}
t_0 = \frac{|\Delta \dot \nu(0)|}{\ddot \nu_{\rm ig}}.
\label{t0}
\end{equation}
With the expectation that the recovery establishes 
conditions ripe for the next glitch, $t_0$ can be used a rough estimate of the time $t_{\rm g}$ to 
the next glitch \citep{alpar84c,akbal17,erbil20}. These rough estimates can be improved by allowing for 
`persistent shifts' in the spin-down rate. First observed systematically in the Crab pulsar's 
glitches \citep{lyne92,lyne15}, a `persistent shift' is a part of the glitch step in the spin-down rate which does not recover at all.
For Crab glitches the values of $ \Delta \dot{\nu}_{\rm p}/ \dot{\nu} $ are of the order 
of $ 10^{-4} $ \citep{wong01,lyne15,ge20}.  
Persistent shifts in the Crab pulsar glitches were interpreted by \citet{alpar96} as due to 
permanent structural changes in the crustal moment of inertia. According to this interpretation,
glitches involve crust cracking in conjunction with vortex unpinning. The offset (persistent shift) 
in the spin-down rate is interpreted as due to creation of new vortex trap regions, which do not participate 
in vortex creep, as a permanent effect of crust cracking \citep{alpar96,erbil19}. 

The relation between the 
inertial moment $I_{\rm b}$ of the vortex trap regions with the persistent shift in $ \dot{\nu} $ is given 
by $ (\Delta \dot{\nu}_{\rm p}/\dot{\nu})=I_{\rm b}/I $. 
$ \dot{\nu}(t)$ is restoring with a constant $ \ddot{\nu}_{\rm ig}$. 
Taking the persistent shift into account, conditions ripe for the next glitch 
are reached at a time 
\begin{equation}
t_{\rm 0, p} = \frac{\left|\Delta \dot \nu(0)-\Delta \dot{\nu}_{\rm p}\right|}{\ddot \nu_{\rm ig}}.
\label{t0p}
\end{equation}
Persistent shifts may be common in all pulsar glitches if the unpinning events 
take place in conjunction with crust breaking and formation of vortex traps, as indicated 
by the model developed for PSR J1119$-$6127 \citep{akbal15} and the Crab pulsar \citep{erbil19}. For the Vela pulsar, which is the most frequent glitcher 
after PSR J0537$-$6910 and the Crab pulsar, estimates of the time to the next glitch improve significantly when 
persistent shifts with $\Delta \dot{\nu}_{\rm p}/ \dot{\nu}\sim 10^{-4}$ are allowed \citep{akbal17}.

\subsection{Model Fits}
We apply timing fits to post-glitch $ \dot{\nu} $ data after 
35 out of the 45 PSR J0537--6910 glitches using the model described above. We did not analyze the remaining 10 glitches  for one of the following reasons: (i) the change in $ \dot{\nu} $ at the glitch is not resolved, (ii) post-glitch datapoints are very sparse, (iii) the $ \dot{\nu} $ measurements have very high error bars, and (iv) the glitch has a negative sign ('anti-glitch'). Early exponential relaxation components are not detectable in the $ \dot{\nu} $ data with its very short inter-glitch time intervals \citep{antonopoulou18}, hence the expressions used for the fits contain only combinations of two relaxation signatures of the non-linear regimes: (i) the full integrated response as given in Eq. (\ref{creepfull}) with three free parameters ($ I_{\rm A}/I $, $ \tau_{\rm nl}$, $ t_{\rm 0} $), (ii) only the approximate integrated response, the linear power law given in Eq. (\ref{glitchdotnu}) with two free parameters ($ I_{\rm A}/I $, $ t_{\rm 0} $), or (iii) the linear power law recovery given in Eq. (\ref{glitchdotnu}) plus the Fermi function form representing the response of a single layer, as given in Eq. (\ref{creepsingle}) with five free parameters ($ I_{\rm A}/I $, $ I_{\rm s}/I $, $ \tau_{\rm nl} $, $ t_{\rm 0} $, $ t_{\rm s} $). In our model $t$ is the time since the glitch date, shifted by the uncertain parameter $ \Delta $ defined as the time lag between the actual glitch date and the first post-glitch observation. 

The best fit values of the fitted and inferred parameters found using the Levendberg-Marquardt procedure \citep{markwardt09} are listed in Table \ref{fitpar} and three examples of the best fit plots are shown in Figure 1. The inferred parameters $ I_{\rm A}/I $, $ I_{\rm s}/I $, $ I_{\rm B}/I $, the total crustal superfluid moment of inertia $ I_{\rm cs}/I $, the offset times $t_{\rm s}$, $t_{0}$ and number of vortices$ N_{\rm step}^{\rm vortex} $, $ N_{\rm int}^{\rm vortex} $  participating in each glitch corresponding to the step [Eq.(\ref{creepsingle})] and integrated [Eqs. (\ref{creepfull}) and (\ref{glitchdotnu})] responses are also tabulated in Table \ref{fitpar}.

\begin{table*}
\label{fitpar}\centering\scriptsize
\caption{The best fit values for the free and inferred parameters for the glitches of PSR J0537--6910. Errors are given in parentheses. We could not analyze 10 glitches for the reasons for a) the change in $ \dot{\nu} $ at the glitch epoch is not resolved, b) post-glitch datapoints are very sparse, c) the $ \dot{\nu} $ measurements have very high error bars, d) the glitch has a negative sign ('anti-glitch').}
   \begin{tabular}{ccccccccccc}
    Glitch Nr. & $ I_{\rm A}/I (10^{-4}) $ & $ I_{\rm s}/I (10^{-4}) $ & $ I_{\rm B}/I (10^{-3}) $ & $ I_{\rm cs}/I (10^{-2}) $  & $ \tau_{\rm nl} $ (days) & $ t_{\rm 0} $ (days) & $ t_{\rm s} $ (days) &$ N_{\rm int}^{\rm vortex}(10^{13}) $ & $ N_{\rm step}^{\rm vortex} (10^{13})$& $ \Delta $ (days)\\
    \hline
    1    & 5.2(9)  & -    & 12.4 & 1.3 & 20(8)  &  196(14) & -  &  6.6 & - & 16\\
    2    & 5.6(8)  & 1.7(6)    & 4 & 0.5  & 4(2) & 301(61)  &  84(3) & 10.2 & 2.9 & 15 \\
    3    & 8(2)  & -    & 15.8 & 1.7  &19(9) & 70(15)  &  - &  2.4 & - & 5 \\
    4    & 5.62(3)  & -    & 7.7 & 0.8  & - & 63(1)  &  -  &2.2 &- & 8\\
    5    & 5.70(5)  & -    & 10.5 & 1.1  &  - & 47(1)  &  - & 1.6& -& 6\\
    6    & 6.7(4)  & 1.8(4)    & 4.9 & 0.6  & 5(3) & 256(17)  &  60(3) & 8.7 & 2.1 & 5\\
    7$ ^{\rm a} $    & -  & -    & - &- & -  &  - & -  &  - & - & -\\
    8    & 7.285(5)  & -    & 5.5 & 0.6  & - & 101(1)  &  - &  3.4 & - & 5\\
    9    & 3.2(1)  & -    & 14.8 & 1.5  & - & 105(4)  &  - & 3.5 & -& 12 \\
    10    & 5.058(7)  & -    & 6.6 & 0.7  & - & 88(1)  &  - & 3.0 & - & 12 \\
    11    & 5(2)  & 1.2(7)    & 6.9 & 0.8  & 12(5) & 72(4)  &  39(7) & 2.5 & 1.3 & 7\\
    12    & 4.5(2)  & -    & 11.2 & 1.2  & - & 133(5)  &  - & 4.5 & - & 7\\
    13    & 5.79(2)  & -    & 2.1 & 0.3  & - & 124(3)  &  - &  4.2 & - & 14\\
    14    & 4.7(3)  & -    & 15.4 & 1.6  & - & 59(3)  &  - &  2.0 & - & 16 \\
    15    & 4.8(2)  & -    & 9.8 & 1.0  & - & 84(2)  &  - & 2.9 & - & 3 \\
    16    & 9(3)  & -    & 8.4 & 0.9  & 30(5) & 138(34)  &  - &  4.7 & - & 6 \\
    17$ ^{\rm b} $    & -  & -    & - & -  & - & -  &  - & - & - & - \\
    18    & 5.04(7)  & -    & 10.1 & 1.1  & - & 136(3)  &  - & 4.6 & - & 2 \\
    19$ ^{\rm c} $    & -  & -    & - & -  & - & -  &  -  & - & - & - \\
    20    & 8.4(3)  & 1.6(2)    & 5.8 & 0.7  & 6(2) & 114(5)  &  38(5) &  3.9 & 1.3 & 2 \\
    21    & 6.1(7)  & 1.8(8)    & 6.1 & 0.7  & 5(1) & 151(12)  &  31(14) & 5.1 & 1.1 & 3 \\
    22    & 6.3(1)  & -    & 7.9 & 0.9  & - & 179(5)  &  - & 6.1 & - & 15\\
    23    & 8.1(2)  & -    & 9.5 & 1.0  & - & 85(18)  &  - & 2.9 & - & 1\\
    24    & 5.517(5)  & -    & 2.4 & 0.3  & - & 25(1)  &  - &  0.8 & - & 2 \\
    25    & 11.150(7)  & -    & 2.4 & 0.4  & - & 124(2)  &  - &  4.2 & - & 4 \\
    26    & 11.2(3)  & -    & 3.1 & 0.4  & - & 265(40)  &  - & 12.4 & - & 5\\
    27$ ^{\rm a} $    & -  & -    & - & -  & - & -  &  - & - & - & - \\
    28    & 8(3)  & -    & 12.3 & 1.3  & 27(15) & 139(55)  &  - & 4.7 & - & 2 \\
    29    & 7.356(4)  & -    & 7.8 & 0.9  & - & 106(1)  &  - & 3.6 & - & 7 \\
    30$ ^{\rm c} $    & -  & -    & - & -  & - & -  &  - & - & - & -\\
    31$ ^{\rm d} $    & -  & -    & - & -  & - & -  &  - &  - & - & - \\
    32$ ^{\rm c} $    & -  & -    & - & -  & - & -  &  - & - & - & - \\
    33    & 10.13(2)  & -    & 9.7 & 1.1  & - & 37(1)  &  - & 1.3 & - & 2 \\
    34    & 11.0(4)  & -    & 5.2 & 0.6  & - & 225(20)  &  - & 7.7& - & 4 \\
    35    & 4.76(6)  & -    & 11.1& 1.2 & - & 115(2)  &  - & 3.9 & - & 9 \\
    36    & 7.9(2)  & -    & 8.7 & 1.0  & - & 86(2)  &  - & 2.9 & - & 2\\
    37    & 10.96(2)  & -    & 1.6 & 0.3  & - & 151(4)  &  -  &5.1& - & 2 \\
    38    & 11.0(3)  & -    & 3.8 & 0.5  & - & 259(24)  &  - & 8.8 & - & 4\\
    39    & 6.00(2)  & -    & 8.3 & 0.9  & - & 71(9)  &  - & 2.4& -  & 7\\
    40    & 3.2(4)  & -    & 3.4 & 0.4  & - & 72(14)  &  - & 2.4 & - & 2\\
    41$ ^{\rm d} $    & -  & -    & - & -  & - & -  &  - & - & - & -\\
    42$ ^{\rm c} $    & -  & -    & - & -  & - & -  &  - & - & -& - \\
    43    & 3.7(2)  & 1.2(3)    & 11.6 & 1.2  & 3(1) & 98(3)  &  41(12) & 3.3 & 1.4 & 4 \\
    44$ ^{\rm d} $    & -  & -    & - & -  & - & -  &  - & - & - & - \\
    45    & 9(3)  & -    & 10.1 & 1.1  & 26(4) & 118(40)  &  - & 4.0 & - & 4\\
   
    \hline
    \end{tabular}
\label{fitpar}
\end{table*}

\begin{figure*}
\centering

\subfloat[Glitch 1]{\includegraphics[width = 3in,height=5.0cm]{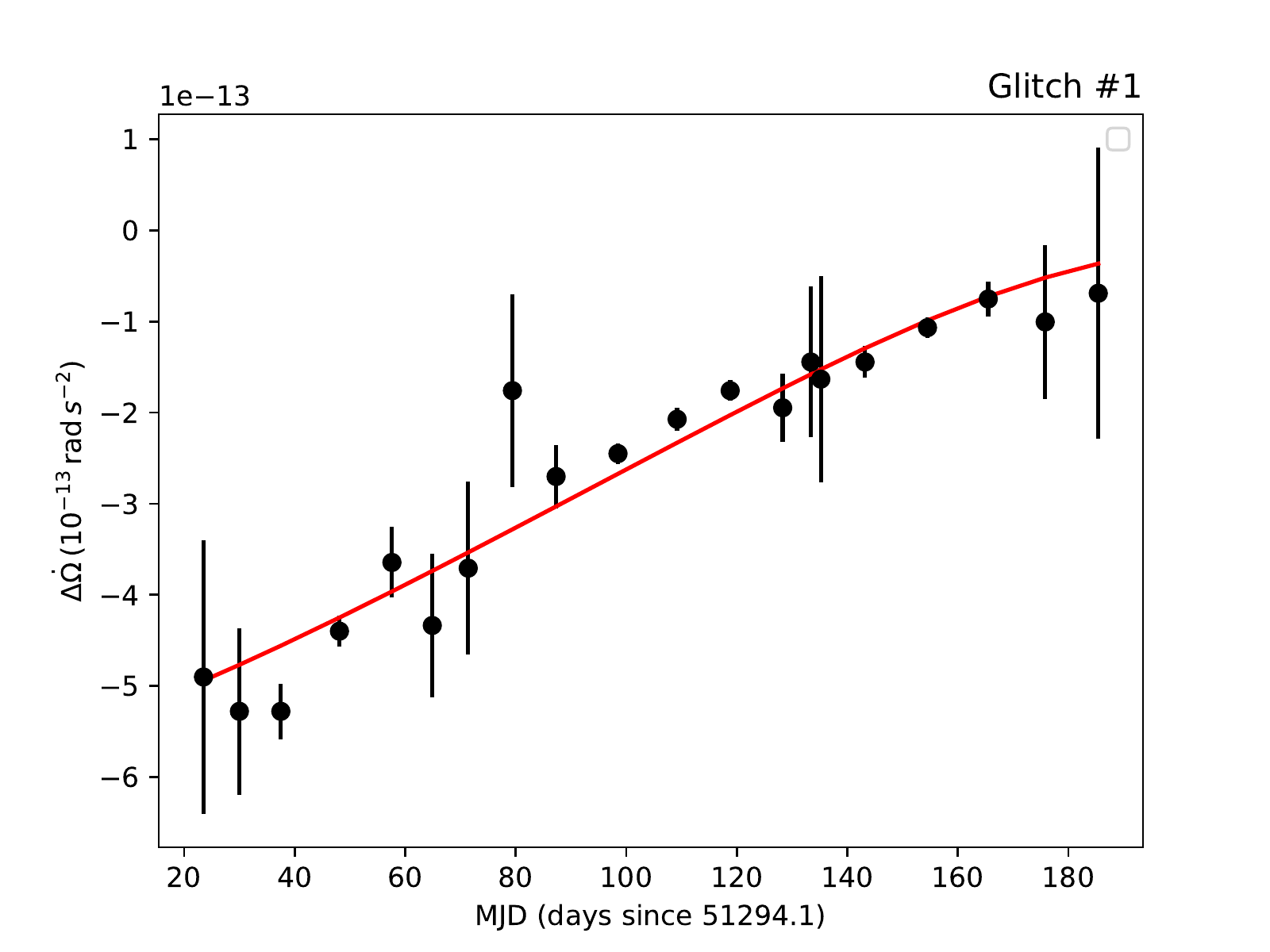}}
\subfloat[Glitch 6]{\includegraphics[width = 3in,height=5.0cm]{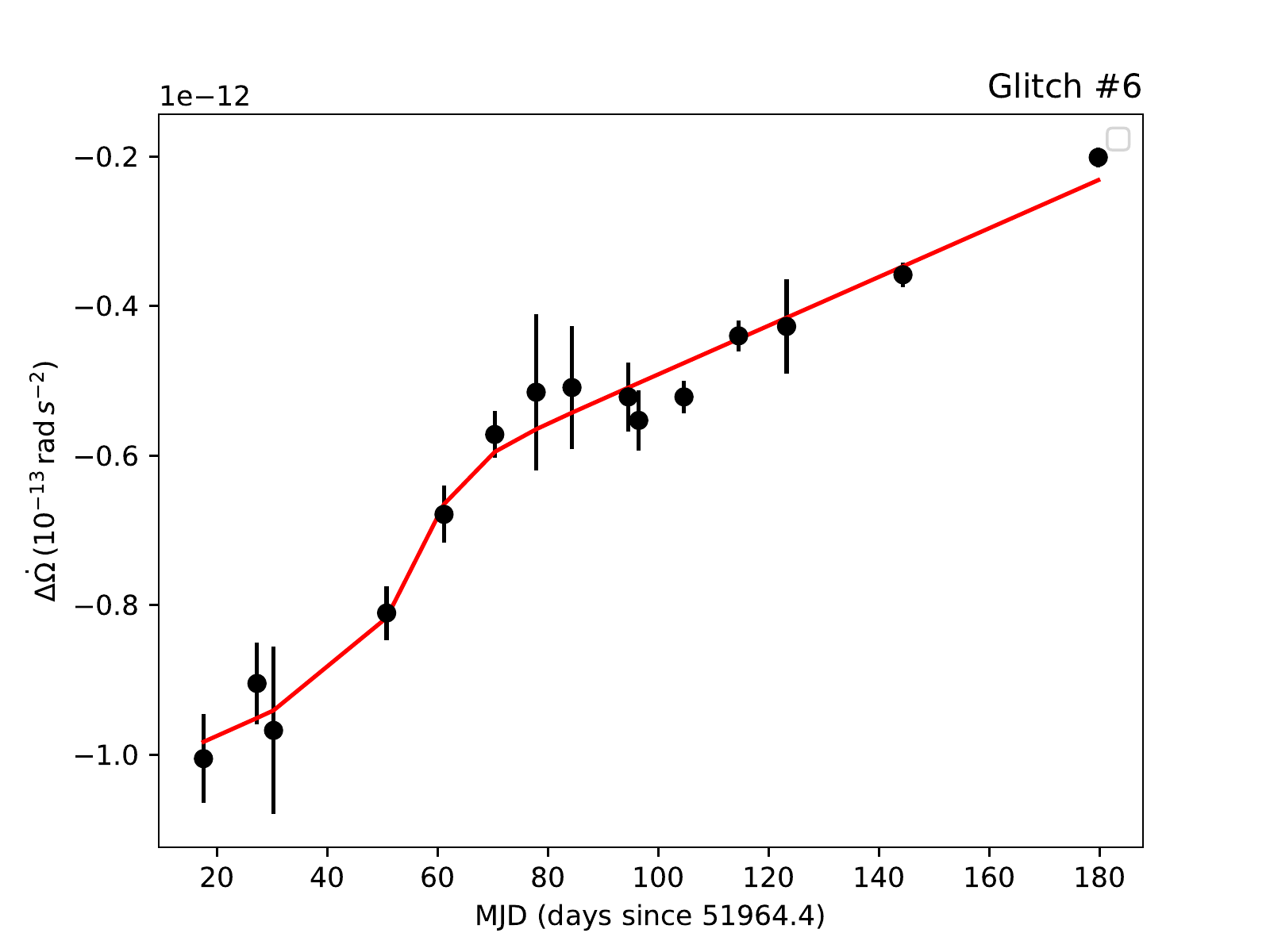}}\
\subfloat[Glitch 35]{\includegraphics[width = 3in,height=5.0cm]{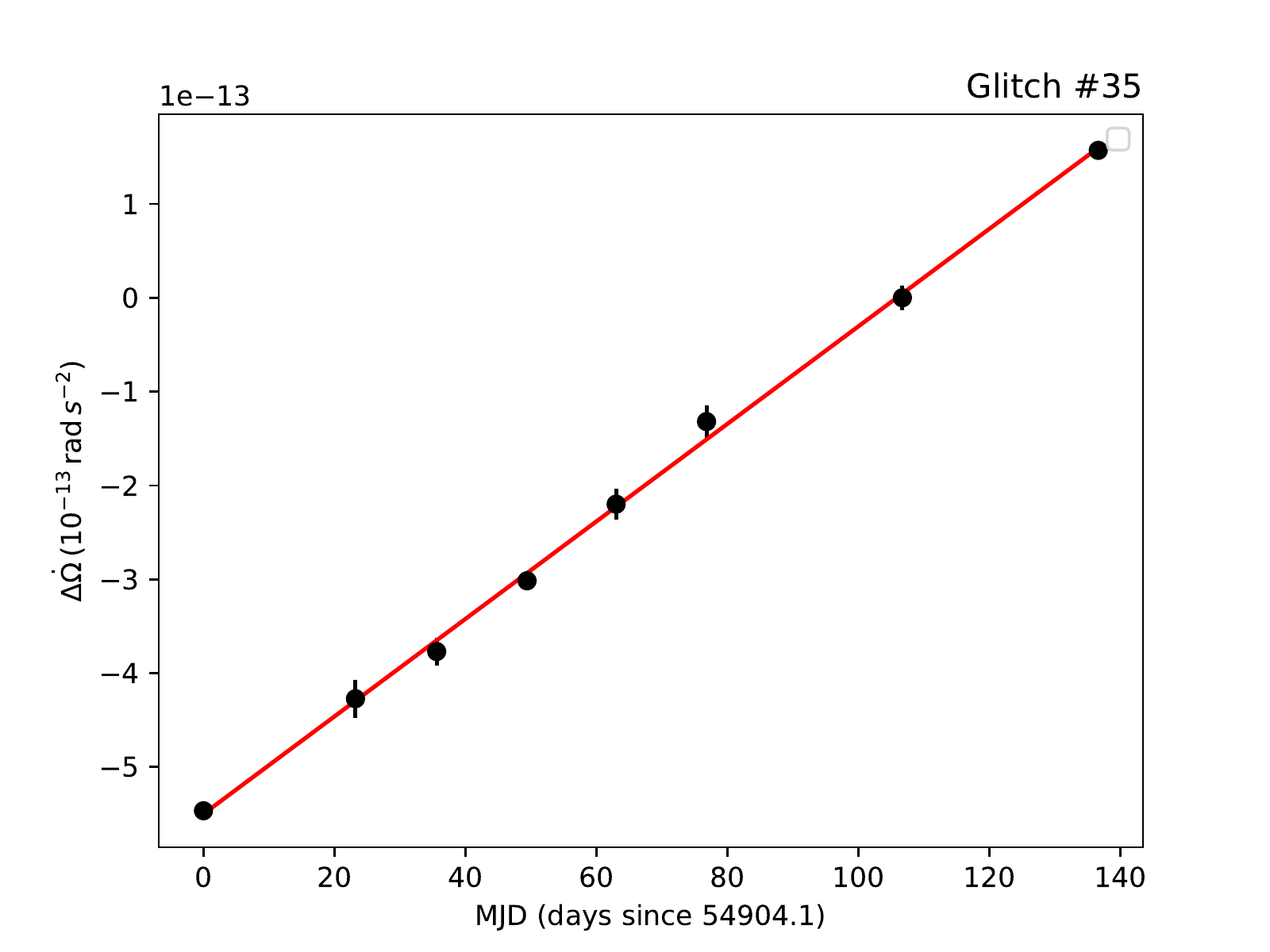}}

\caption{Model fits to the post-glitch spin-down rate after the 1st, 6th, and 35th glitches of PSR J0537$-$6910. Fit plots for the remaining glitches can be found in the supplementary material.}
\label{some example}
\end{figure*}

\section{Braking Index Determination}
\label{sec:braking}

There is considerable ambiguity regarding the rotational braking of PSR J0537$-$6910, in particular about the determination of its actual braking index. For a pulsar with plasma filled magnetoshere the braking index is expected to lie in the range $3\leq n\leq3.25$ under the central dipole approximation \citep{arzamasskiy15, eksi16}. In contrast to other pulsars, the spin-down rate of PSR J0537$-$6910 is increasing in the long-term which indicates a negative long term second derivative of frequency and in turn a negative braking index. Since the rotational evolution of PSR J0537$-$6910 is frequently interrupted by large glitches with magnitude $\Delta\nu/\nu=\mbox{a few times}\times10^{-7}$ in every $\sim100$ days, glitch contributions should largely affect its rotational parameters. In order to obtain the true rotational parameters for PSR J0537$-$6910 these glitch contributions should be assessed and removed from the data. Before proceeding to a discussion of how to tackle this issue it will be useful to discuss the existing approaches in the literatute leading to different inferences for the braking index of PSR J0537$-$6910:
\begin{enumerate}
\item \citet{ferdman18} employ a method in which the first observed post-glitch value of the spin-down rate is subtracted from the whole post-glitch data string for each glitch. They then plot the general trend obtained versus time since the glitch for 42 glitches [see Figure 7 in \citet{ferdman18}]. Their analysis using Monte Carlo simulations included an exponential decay of the initial increase of the post glitch spin-down rate with a time-scale of $\tau_{\rm d}=27^{+7}_{-6}$ days followed by a recovery of the remaining spin-down increment linearly with time. Following the procedure outlined in \citet{espinoza17}, they assumed that the glitch recovery is completed after the exponential decay is over and obtained $\ddot\nu=(4.7\pm0.5)\times10^{-22}$ Hz~\mbox{s$^{-2}$} from the slope of the linear decay trend of the inter-glitch spin-down rate evolution which in turn leads to an unusually high braking index of $n=7.4\pm0.8$. The methodology used by \citet{ferdman18} is misleading in two ways. Firstly, since the glitch induced contributions are relative to the pre-glitch rotational parameters, subtraction of the value of the first post-glitch datapoint from the remaining post-glitch data is irrelevant. Secondly, the response of interior superfluid components to a glitch involves long term linear recovery in time as well as the prompt exponential decay. The resulting inter-glitch $\ddot\nu_{\rm ig}$ differs largely from $\ddot\nu_{0}$ corresponding to the stellar spin-down under external braking, leading  instead to large ‘anomalous’ braking indices reflecting internal torques of superfluid response [cf. equations (\ref{glitchdotnu}) and (\ref{ddotnuig})]. Thus, the braking index derived by \citet{ferdman18} does not provide an estimate of the `true' braking index due to the external pulsar torque.  
\item \citet{antonopoulou18} quantified the large inter-glitch braking indices for different data segments in the range $n_{\rm ig}\simeq8-100$ for 45 glitches observed in  PSR J0537$-$6910. To evaluate the long term braking index they picked the last post-glitch data point before the succeeding glitch and constructed the long-term evolution of the spin-down rate by joining these points. Then, from the slope of the resulting linear trend they obtained $\ddot\nu=-7.7(3)\times10^{-22}$ Hz~\mbox{s$^{-2}$} which leads to a negative braking index, $n=-1.22(4)$. \citet{ho20} have analysed 8 more glitches of PSR J0537$-$6910. Repeating the method described in \citet{antonopoulou18} they find a similar braking index $n=-1.25\pm0.01$. This approach implicitly assumes that the glitch step in spin-down rate is completely recovered by the post-glitch relaxation by the time the next glitch arrives. As will be shown below this is not the case: unresolved persistent (i.e. non-relaxing) parts of the glitch steps in the spin-down rate cumulatively lead to the inference of a long-term negative braking index for PSR J0537$-$6910.
\end{enumerate} 

To find out the underlying braking index due to the pulsar torque let us first define the spin-down rate evolution with long-term glitch induced transient and persistent contributions:
\begin{equation}
\dot\nu(t)=\dot\nu_{0}+\ddot\nu_{0}t+ \sum_{i=1}^{n}\left(\Delta\dot\nu_{\rm i}+\ddot\nu_{\rm ig,i}(t-t_{\rm i})\right)\theta(t-t_{\rm i})\theta(t_{i+1}-t)+\sum_{i=1}^{n}\Delta\dot\nu_{\rm p,i}\theta(t-t_{i}),
\label{nudotev}
\end{equation}
where $\dot\nu_{0}$ and $\ddot\nu_{0}$ are the spin-down rate and second time derivative of the frequency due to external braking torque for a fiducial time, $\Delta\dot\nu_{\rm i}$ and $\ddot\nu_{\rm ig,i}$ are the step increase in the spin-down rate at the \textit{i}th glitch and the observed interglitch second derivative during the recovery which lasts for an interval $ t_{g,i} \equiv t_{i+1}-t_{i}$ until the next glitch. In the observations the behaviour with constant second derivative $ \ddot{\nu}_{g,i} $ continues to the last data point prior to glitch $ i+1 $, so we extrapolate to $ t_{i+1} $ over the usually few days gaps without timing data. $\Delta\dot\nu_{\rm p,i}$ are the step increases which do not relax at all. After a time span including many glitches occurring at times $t_{\rm i}$, equation (\ref{nudotev}) reduces to
\begin{equation}
\dot\nu(t)=\dot\nu_{0}+\ddot\nu_{0}t+\sum_{i=1}^{n}\Delta\dot\nu_{\rm p,i}\theta(t-t_{\rm i}).
\end{equation}
This is because the first sum in equation (\ref{nudotev}), which describes the recovering part of the glitch induced steps in the spin-down rate, is zero after each interglitch recovery lasting for $t_{\rm g,i}$ corresponding to the \textit{i}th glitch. 
Thus, it is the negative persistent shifts $\Delta\dot\nu_{\rm p,i}$ which result in the negative overall slope for $\dot\nu(t)$ despite a positive $\ddot\nu_0 (t)$ from the external (pulsar) torque.

In accordance with equations (\ref{ddotnuig}) and (\ref{nudotev}) we can express the inter-glitch $\ddot\nu_{\rm ig}$ as
\begin{equation}
\ddot\nu_{\rm ig,i}=\ddot\nu_{0}+\frac{I_{\rm A,i}}{I}\frac{\dot\nu_{\rm 0}(t_{\rm i})}{t_{\rm 0,i}}-\frac{\Delta\dot\nu_{\rm p,i}}{t_{\rm 0,i}},
\label{ddotnuigi}
\end{equation}
for the \textit{i}th glitch. An inspection of Table \ref{fitpar} reveals that $t_{\rm 0,i}\approx t_{\rm g,i}\simeq100$ days and since in such a short time interval $\dot\nu_{0}$ does not change appreciably we estimate $\Delta\dot\nu_{\rm p,i}$ as 
\begin{equation}
\Delta\dot\nu_{\rm p,i}=\dot\nu_{\rm 0}(t_{\rm i+1})-\dot\nu_{\rm 0}(t_{\rm i}),
\label{dotnustep}
\end{equation}
as the creep process, stopped by the glitch induced offset, recovers completely around time $t_{\rm 0,i}$. Persistent shifts in the spin-down rate are clearly observed with a magnitude $\Delta\dot\nu_{\rm p}/\dot\nu_{0}\approx4\times10^{-4}$ after the Crab pulsar's glitches \citep{wong01,lyne15,ge20}. Such persistent shifts in the spin-down rate are atributed to the formation of new vortex traps in the crust of young neutron stars as a result of quakes \citep{alpar96,erbil19}. Being a young pulsar like the Crab we expect PSR J0537$-$6910 to display persistent shifts also. Upon applying equations (\ref{ddotnuigi}) and (\ref{dotnustep}) to the observational data and using the corresponding model parameters from Table \ref{fitpar} we obtain the average value of $\langle\Delta\dot\nu_{\rm p}/\dot\nu_{0}\rangle=1.3\times10^{-4}$, which is very close to that of the Crab pulsar, $\langle\Delta\dot\nu_{\rm p}/\dot\nu_{0}\rangle=2.5\times10^{-4}$ \citep{lyne15}. Finally we obtain $\ddot\nu_{0}=1.73(24)\times10^{-21}$ Hz~\mbox{s$^{-2}$} due to spin-down under the external braking torque. This leads to a braking index of $n=2.7(3)$ for PSR J0537$-$6910 after glitch induced transients and persistent contributions to the rotational parameters have beeen removed. This value is consistent with the other braking index measurements from young pulsars and close to that of the next most prolific glitcher, the Vela pulsar, for which $n =2.81(12)$\citep{akbal17}. We also determine the braking index model-independently. Exchanging $ I_{A,i}/I $ and $ t_{0,i} $ in Eq. (14) and Table \ref{fitpar} with the observed parameters \citep{antonopoulou18}, i.e. $ \Delta \dot{\nu} / \dot{\nu} $ and $ t_{g,i} $, we obtain $\ddot\nu_{0}=1.38(17)\times10^{-21}$ Hz~\mbox{s$^{-2}$}. This gives the braking index $ n=2.2(3) $ which is close to the model-dependent braking index value within the uncertainties.

\section{Results}
\label{sec:results}

In this section we infer physical parameters of the neutron star interior for PSR J0537$-$6910 from model fits.  

The non-linear creep relaxation time-scale is given by \citep{alpar84}
\begin{equation}
\tau_{\rm nl}\equiv \frac{kT}{E_{\rm p}}\frac{\omega_{\rm cr}}{\vert\dot{\Omega}\vert},
\label{taunl}
\end{equation}
where $\omega_{\rm cr}$ is the critical lag for unpinning and $E_{\rm p}$ is the pinning energy for the vortex line-lattice nucleus interaction. For a neutron star with a radius $R=12$ km and mass $M=1.6M_{\odot}$ the interior temperature as a result of neutrino cooling via modified URCA reactions is given by \citep{yakovlev11}
\begin{equation}
T_{\rm in}= 1.78\times10^{8}~{\rm K}~\left(\frac{10^{4}{\rm yr}}{t_{\rm sd}}\right)^{1/6},
\label{murca}
\end{equation}
where $t_{\rm sd}=\nu/2\dot\nu$ is the characteristic (spin-down) age of the pulsar. \citet{erbil20} considered five crustal superfluid layers and estimated the range $\omega_{\rm cr}/E_{\rm p}\simeq0.5-0.01$\mevs,~the lowest value being reached in the densest pinning region close to the crust-core interface. Eqs.(\ref{taunl}) and (\ref{murca}) yield the estimate for the nonlinear creep time-scale
\begin{equation}
\tau_{\rm nl}\cong(30-1420)\left(\frac{\dot\nu}{10^{-11}\mbox{\hzs}}\right)^{-1}\left(\frac{t_{\rm sd}}{10^{4}{\rm yr}}\right)^{-1/6}\mbox{days},
\label{creepnl}
\end{equation}
which is monotonically decreasing with increasing density towards the crust-core boundary. With $t_{\rm sd}=4.93$ kyr and $\dot\nu=1.992\times10^{-10}$ \mbox{Hz s$^{-1}$} equation (\ref{creepnl}) gives $\tau_{\rm nl}=(1.7-80)$ days for PSR J0537$-$6910, which is in qualitative agreement with the range obtained from fits to the data, $\tau_{\rm nl}=(3-30)$ days. Alternatively, one can estimate the crustal temperature by using equation (\ref{taunl}) with relevant physical parameters pertaining to the glitch trigger regions of the crust and the recoupling time-scales deduced from fits to the data. 

If we assume that the large glitches of the Vela pulsar start at the innermost and densest region of the neutron star crust, while the smaller glitches of PSR J0537--6910 are triggered at lower densities, we can estimate the density at the glitch trigger region $\rho_{\rm loc}$ in PSR J0537--6910 by comparing the $I_{\rm A}/I$ values inferred for the two pulsars:
\begin{equation}
\frac{\left(\rho_{\rm loc}\right)_{[{\rm PSR~J0537}]}}{\left(\rho_{\rm loc}\right)_{[{\rm Vela}]}}\simeq \frac{\left(I_{\rm A}/I\right)_{[{\rm PSR~J0537}]}}{\left(I_{\rm A}/I\right)_{[{\rm Vela}]}}.
\end{equation}
The mean value of $I_{\rm A}/I$ is $6.1\times10^{-3}$ for the Vela pulsar glitches \citep{akbal17,erbil20} while $\langle I_{\rm A}/I\rangle = 6.9\times10^{-4}$ for PSR J0537$-$6910. This leads to $\rho_{\rm loc}(\mbox{PSR~J0537)}\cong0.11\rho_{\rm loc}(\mbox{Vela})\sim1.5\times10^{13}$ \mbox{g~cm$^{-3}$}. From Table \ref{fitpar} $\langle\tau_{\rm nl}\rangle=14$ days and we arrive at the estimate $T=0.9\times10^{8}$ K for the inner crust temperature of PSR J0537$-$6910.  

\begin{figure}
	\includegraphics[width=\columnwidth]{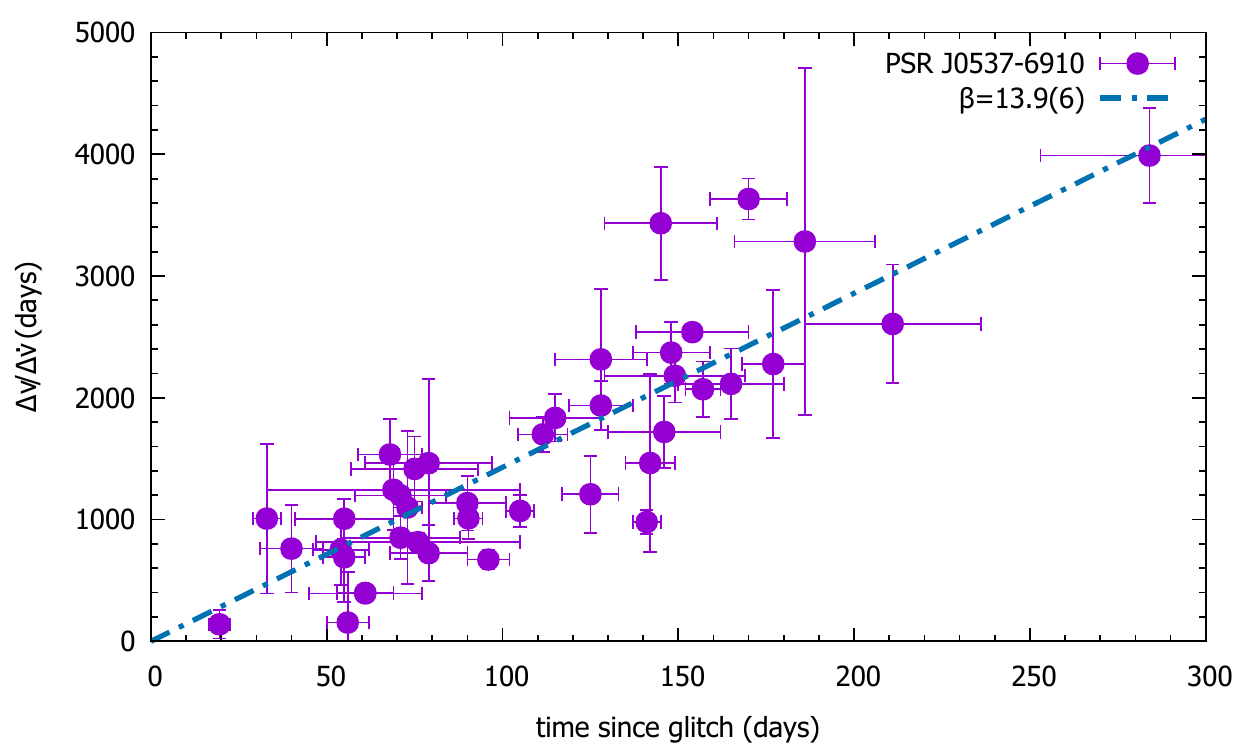}
    \caption {$\Delta\nu/\Delta\dot\nu$ vs. $t_{\rm g}$ relation for PSR J0537$-$6910 glitches. Observational data \citep{antonopoulou18,ferdman18,ho20,abbott20} is shown with (purple) dots. The best fit with equation~(\ref{creepcor}) is shown with the blue dot-dashed line, for $\beta=13.9(6)$.}
    \label{corfit}
\end{figure}

According to the vortex creep model a correlation between the ratio of glitch magnitudes in frequency and spin-down rate $\Delta\nu/\Delta\dot\nu$ and inter-glitch time $t_{\rm g}$ is expected \citep{alpar84}
\begin{equation}
\frac{\Delta\nu}{\Delta\dot\nu}=\left(\frac{1}{2}+\beta\right)t_{\rm g},
\label{creepcor}
\end{equation}
from equations (\ref{glitchdotnu}) and (\ref{glitchomega}). The correlation coefficient gives a measure of $\beta=I_{\rm B}/I_{\rm A}$ which presents the extend of the vortex traps within the neutron star crust. For the data sample containing 41 glitches the best fit linear relation of equation~(\ref{creepcor}) is shown in Figure~\ref{corfit}.
The dashed blue line corresponds to the best fit with equation (\ref{creepcor}) which gives $\beta\equiv I_{\rm B}/I_{\rm A}=13.9(6)$. This ratio is very close the mean value $\langle I_{\rm B}/I_{\rm A}\rangle=13.8$ from Table \ref{fitpar} obtained by fits to the individual inter-glitch timing data. $\langle I_{\rm B}/I_{\rm A}\rangle$ is 2.2 for the Vela pulsar \citep{akbal17,erbil20} while $\lesssim0.25$ for the Crab pulsar \citep{alpar96,erbil19}. It seems for PSR J0537$-$6910 that $I_{\rm A}/I$ values are similar to those of the Crab pulsar while  $I_{\rm B}/I$ values are close to those of the Vela pulsar. The relative smallness of  $I_{\rm A}/I$ means either that the number of vortex traps is low or that they are not interconnected. On the other hand, large $I_{\rm B}/I$ means that unpinned vortices scatter through large vortex free regions before they can repin. 

The number of the vortex lines unpinned in a glitch can be related to the broken plate size $D$ through \citep{akbal15}
\begin{equation}
\delta N_{\rm v}=2\pi R^{2}\frac{\delta\Omega_{\rm s}}{\kappa}=2\pi R\sin{\alpha}D\frac{2\Omega}{\kappa}\sim2\pi RD\frac{2\Omega}{\kappa},
\label{vornum}
\end{equation}
provided the angle $\alpha$ between the position of the plate and the rotation axis is not small. Here $\kappa=h/2m_{\rm n}=2\times10^{-3}$\cmss ~is the vortex quantum. For PSR J0537$-$6910 $\Omega\cong390$ \mbox{rad~s$^{-1}$} and from Table \ref{fitpar} the average number of vortex lines unpinned at the time of the glitch is $\langle\delta N_{\rm v}\rangle=4.65\times10^{13}$. Equation (\ref{vornum}) then yields the estimate $D=19$ cm for the broken plate size. This is smaller than the previous estimates of the broken plate size, $D \sim$ 6 m for PSR J1119-6127 \citep{akbal15}, $D \sim6-18$ m for the Crab pulsar \citep{erbil19}. The clue to this discrepancy may be the fact that PSR J0537$-$6910 is not a radio pulsar despite its rapid rotation rate. This would mean that the angle $\alpha$  between the rotation and  magnetic axes is small. If magnetic stresses play some role in crust breaking, the plate would be close to the magnetic pole. In this case the distance from the rotation rate, $R \sin{\alpha} \sim R \alpha \ll R$, and the plate size $D$ would be much larger than 20 cm, and could well be in the $\sim 10$ m range inferred for the other pulsars. The fact that $\delta N_{\rm v}\sim10^{13}-10^{14}$ is constant within a factor of a few among pulsars of different ages which exhibit glitches of various magnitudes can be understood as a natural outcome of the idea of universality of the plate size broken in a quake as a trigger for the glitch, determined by the critical strain angle $\theta_{\rm cr}$ in the neutron star crust solid \citep{akbal15,akbal18,erbil19}. The critical strain angle $\theta_{\rm cr}$ at which the solid crust yields is related to the plate size and crust thickness $\ell_{\rm crust}$; $\theta_{\rm cr}\sim D/\ell_{\rm crust}$ with $\ell_{\rm crust}\approx0.1R\sim10^{5}$ cm. From theoretical and numerical calculations the range $\theta_{\rm cr}\sim 10^{-2}-10^{-1}$ was obtained for the unscreened Coulomb solid in the neutron star crust \citep{baiko18}.

Glitch statistics reveal that the ratio $\delta\Omega_{\rm s}/\Omega$ does not deviate significantly from glitch to glitch among pulsars \citep{alpar94} which reflects the finite extent of the angular momentum reservoir depleted at each glitch. On taking the average value for $\delta\Omega_{\rm s}/\Omega$ from the analysis of the Vela glitches by \citet{alpar94} an estimate for the waiting time, Eq.(\ref{waitingtime}), between large glitches can be recast in terms of the spin-down age of a pulsar \citep{alpar06}:
\begin{equation}
t_{\rm ig}=\frac{\delta\Omega_{\rm s}}{\Omega}\frac{\Omega}{\left|\dot\Omega\right|}\cong2\left<\frac{\delta\Omega_{\rm s}}{\Omega}\right>t_{\rm sd}\cong3.5\times10^{-4}t_{\rm sd}.
\label{gtime}
\end{equation}
With $t_{\rm sd}=4.93$ kyr equation (\ref{gtime}) predicts $t_{\rm ig}=630$ days for PSR J0537$-$6910 which is about five times larger than the observed inter-glitch timescale $t_{\rm ig, obs}=120$ days. This may be because PSR J0537$-$6910 is not a Vela-like pulsar and is still in the process of developing new vortex traps. 

An exponential relaxation is resolved following some of the PSR J0537--6910 glitches. After the first glitch in the RXTE data \citet{antonopoulou18} identified a exponentially decaying component with a time-scale $\tau_{\rm d}=21(4)$ days. For the second and fourth glitch in the NICER data \citet{ho20} found marginal evidence of a exponential recovery with $\tau_{\rm d}\sim 5$ days. These results are in line with estimates for the exponential relaxation time of vortex creep across toroidal flux lines of the proton superconductor in the outer core of the neutron star \citep{erbil17}. The model predicts $\tau_{\rm tor}=(1-20)$ days for the exponential decay time-scale for PSR J0537--6910 parameters. From these exponential decay time-scales we estimate the pinning energy per vortex-flux tube junction for the core superfluid of PSR J0537--6910 to be $E_{\rm v-\Phi}=4.32$ MeV \citep{erbil16}. The radial extent of the toroidal field region in terms of the fractional moment of inertia can be inferred from post-glitch data fits for the amplitude of the exponentially decaying component and the total glitch magnitude as $I_{\rm tor}/I=\Delta\nu_{\rm d}/\Delta\nu_{\rm g}$ \citep{erbil17}. The parameters of the first glitch, $\Delta\nu_{\rm d}=0.3(1)\mu$Hz and $\Delta\nu_{\rm g}=42.3(2)\mu$Hz \citep{antonopoulou18} yield  $I_{\rm tor}/I=(7\pm2)\times10^{-3}$ for PSR J0537--6910.  

\section{Discussion and Conclusions}
\label{sec:disandconc}

In this paper we have studied the long term timing behaviour of the unique source PSR J0537--6910. Two characteristics make this pulsar special and worthy of detailed investigation \citep{antonopoulou18,ferdman18}:
\begin{enumerate}
\item Spin evolution is frequently interrupted by glitches of magnitude $\Delta\nu/\nu\gtrsim10^{-7}$ every $\sim100$ days in a way such that the glitch magnitude and the time to the next glitch are interrelated. 
\item These glitches leave behind persistent, non-relaxing remnant steps in the spin-down rate so that there is an apparent linear increase trend in the long term spin-down rate leading to an apparent negative braking index in contrast to the spin evolutions of other pulsars. 
\end{enumerate}
Close examination of item (i) reveals the mechanisms responsible for frequent discrete angular momentum transfer and its relaxation associated with glitches as well as constraining physical parameters of the neutron star crust and core. On the other hand item (ii) allows us to assess transient and persistent contributions to the rotational parameters arising from glitches and inter-glitch relaxation associated with the internal dynamics of the neutron star. Extracting the internal dynamics allows us to identify the true pulsar braking index associated with the external torque with some confidence.     

In relation to (i) we reanalysed the first 45 glitches of PSR J0537$-$6910 published in the literature. We have done post-glitch timing fits within the vortex creep model. From our post-glitch timing fits we determined moments of inertia of the superfluid regions involved in glitches as well as recoupling time-scales of the glitch affected superfluid regions as given in Table \ref{fitpar}. We argued that all of these values are consistent with theoretical expectations in that a young and hot neutron star like PSR J0537--6910 should resume steady state behaviour in a short time after glitches and the superfluid regions responsible for collective vortex unpinning may not involve the whole crust \citep{alpar89,alpar96}. 

We have identified in the PSR J0537-6910 timing data a fundamental correlation, familiar, in particular, from the analyses of the Vela pulsar \citep{alpar84c},  between the ratios of the glitch magnitudes in rotation and spin-down rates and the time to the next glitch, i.e. $\Delta\nu/\Delta\dot\nu\propto t_{\rm g}$  (correlation coefficient 0.95), see Figure (\ref{corfit}). This is an expected relation in terms of the vortex creep model with the proportionality factor given in equation (\ref{creepcor}) involving the ratio of the moment of inertia of the vortex trap regions $I_{\rm B}$ to the moment of inertia of the nonlinear creep regions $I_{\rm A}$ giving rise to the vortex unpinning avalanche. We observe that $I_{\rm A}$ values for PSR J0537--6910 are close to those of the Crab pulsar while $I_{\rm B}$ values for PSR J0537--6910 are like those of the Vela pulsar. Since the age of PSR J0537--6910 is in between the ages of the Crab and Vela pulsars, such a connection is in line with a scenario of evolution of glitch properties. That is, when a pulsar is younger and hotter the number and the extent of the nonlinear creep regions involved in the vortex unpinning avalanche are not large while vortices scatter to greater distances as a pulsar cools and more vortex traps with vortex free scattering regions are formed \citep{erbil20}. We estimated the inner crust temperature of PSR J0537--6910 as $T=0.9\times10^{8}$ K from microphysical parameters of the crust and $\tau_{\rm nl}$ values from model fits. Notice that $\left[T_{\rm PSR~J0537}/T_{\rm Vela}\right]=\left(t_{\rm sd, Vela}/t_{\rm sd, PSR~J0537}\right)^{1/6}$ \citep{erbil20}, so we infer that the Vela and  PSR J0537--6910 obey the same cooling behaviour. 

If the broken plate is in the magnetic polar cap region which is close to the rotation axis as suggested by the absence of radio emission activity for PSR J0537--6910, the broken plate size could be $\sim10$ m as previously inferred for PSR J1119--6127 \citep{akbal15} and the Crab pulsar \citep{erbil19}. The size of the broken crustal platelets seem to be a universal structural property of a neutron star reflecting the critical strain angle of the Coulombic crust solid. 

From the short term post-glitch relaxation of  PSR J0537--6910 we determined the volume occupied by toroidal field lines in the core as $I_{\rm tor}/I=(7\pm2)\times10^{-3}$ and pinning energy per vortex-flux tube intersection as $E_{\rm v-\Phi}=4.32$ MeV.    

In relation to (ii) we detailed and employed a method for finding true braking indices of pulsars by removing transient and persistent contributions of glitches to the rotational parameters. We have shown that glitches of PSR J0537--6910 introduce persistent increases to the spin-down rate on average by an amount $\Delta\dot\nu_{\rm p}/\dot\nu\approx1.5\times10^{-4}$ similar to the case of the Crab glitches. By cleaning these increments we are able to obtain a braking index $ n=2.7(3) $ for PSR J0537--6910. Note that to interpret the large interglitch braking indices \citet{antonopoulou18} and \citet{andersson18} pursued the idea of possible emission of gravitational waves associated with PSR J0537--6910 glitches. Continuous gravitational radiation of neutron stars at once or twice the rotational frequency might arise due to crustal deformation \citep{cutler02} and internal fluid motions arising from r-modes \citep{andersson18} leading to braking indices $n=5$ and $n=7$, respectively. Initial searches with LIGO/Virgo third survey failed to detect  gravitational radiation associated with PSR J0537--6910 due to either mechanisim \citep{fesik20,abbott20,abbott21}. If our deduced value of $ n=2.7(3) $, extracted by ascribing the interglitch behaviour to internal dynamics of the neutron star, is close to the actual braking index for PSR J0537--6910, then this pulsar will not be a source for continuous gravitational wave radiation. 

\section*{Acknowledgements}
We thank Danai Antonopoulou and Cristobal Espinoza for sharing the digital data published in \citet{antonopoulou18} with us. We acknowledge the Scientific and Technological Research Council of Turkey
(T\"{U}B\.{I}TAK) for support under the grant 117F330.
\section*{Data Availability}
No new data were analysed in support of this paper.


\bibliographystyle{mnras}



\appendix

\section{Supplementary Material}

We analyze 35 out of the 45 post-glitch $ \dot{\nu} $ relaxations of PSR J0537--6910.  We did not analyze the remaining 10 glitches  for one of the following reasons: (i) the change in $ \dot{\nu} $ at glitch epoch is not resolved, (ii) post-glitch datapoints are very sparse, (iii) the $ \dot{\nu} $ measurements have very high error bars, and (iv) the glitch has a negative sign ('anti-glitch'). Early exponential relaxation components are not detectable in the $ \dot{\nu} $ data due to its very short inter-glitch time intervals. The expressions used for the fits contain only the combinations of two relaxation signatures of the non-linear regimes: (i) only integrated fermi-like function for 5 glitches, (ii) only power law function for 24 glitches, and (iii) power law recovery plus fermi-like function responses for 6 glitches.

\begin{figure*}
\centering

\subfloat[Glitch 1]{\includegraphics[width = 3in,height=5.0cm]{G1.pdf}}
\subfloat[Glitch 2]{\includegraphics[width = 3in,height=5.0cm]{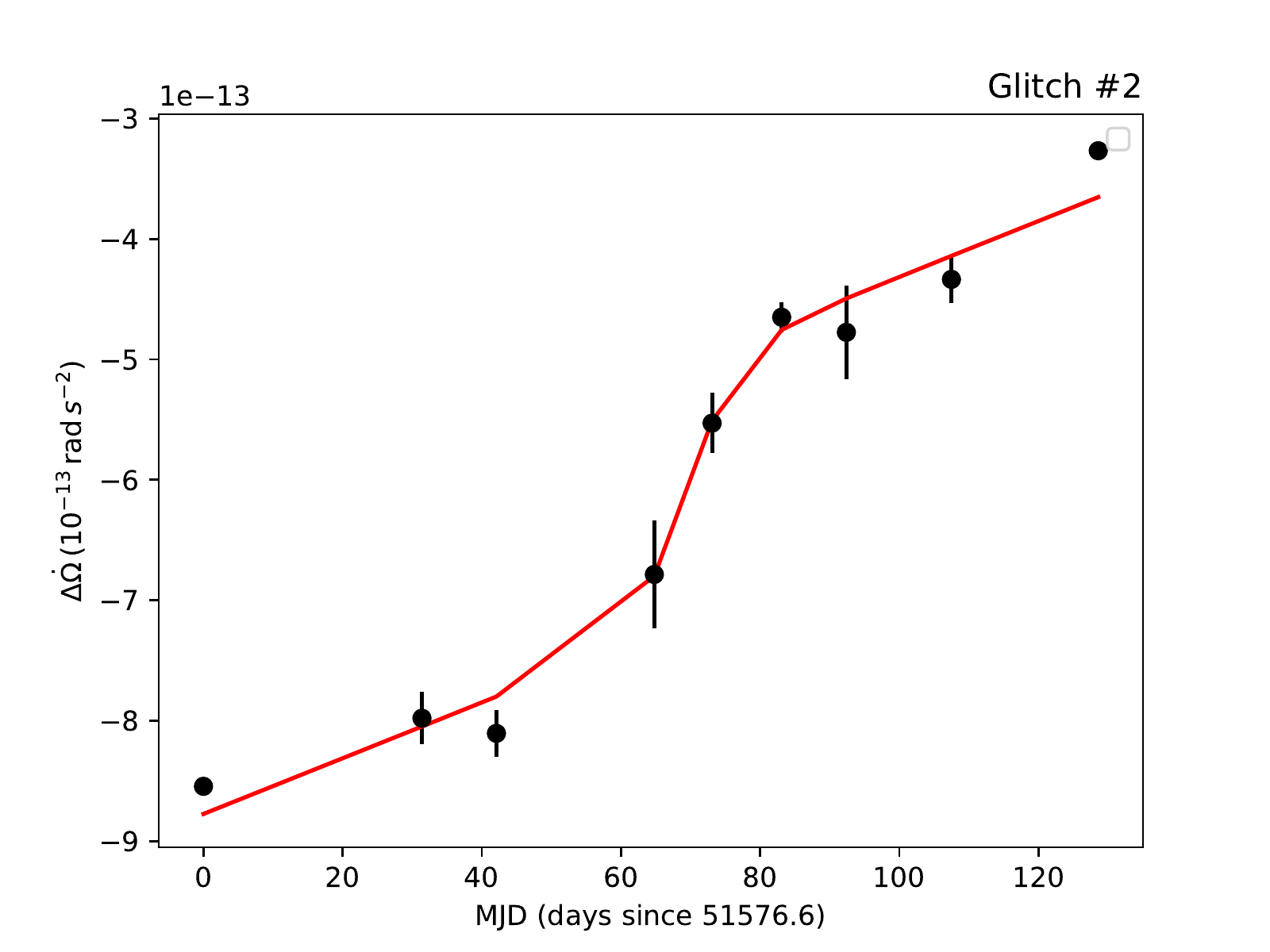}}\
\subfloat[Glitch 3]{\includegraphics[width = 3in,height=5.0cm]{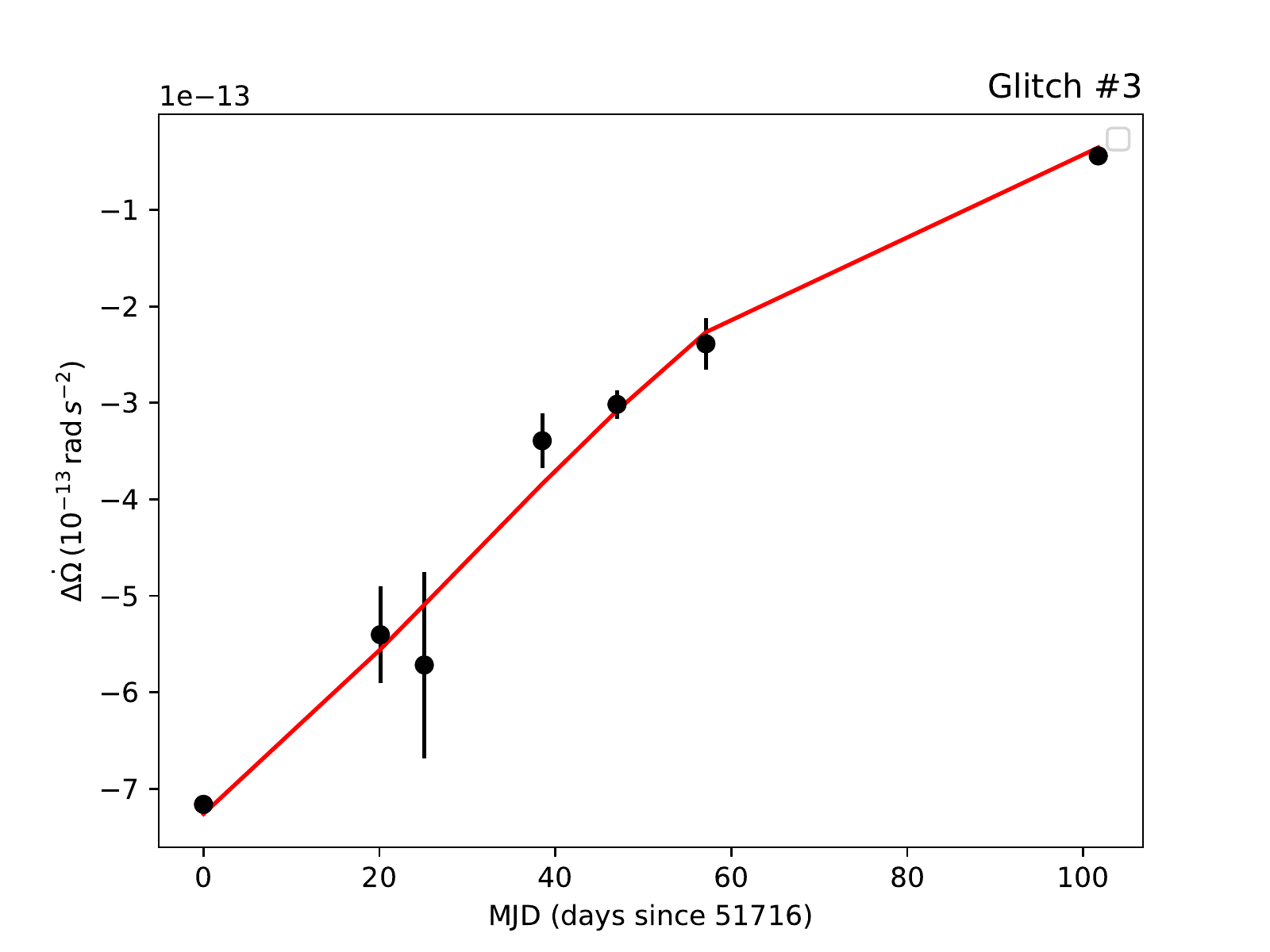}}
\subfloat[Glitch 4]{\includegraphics[width = 3in,height=5.0cm]{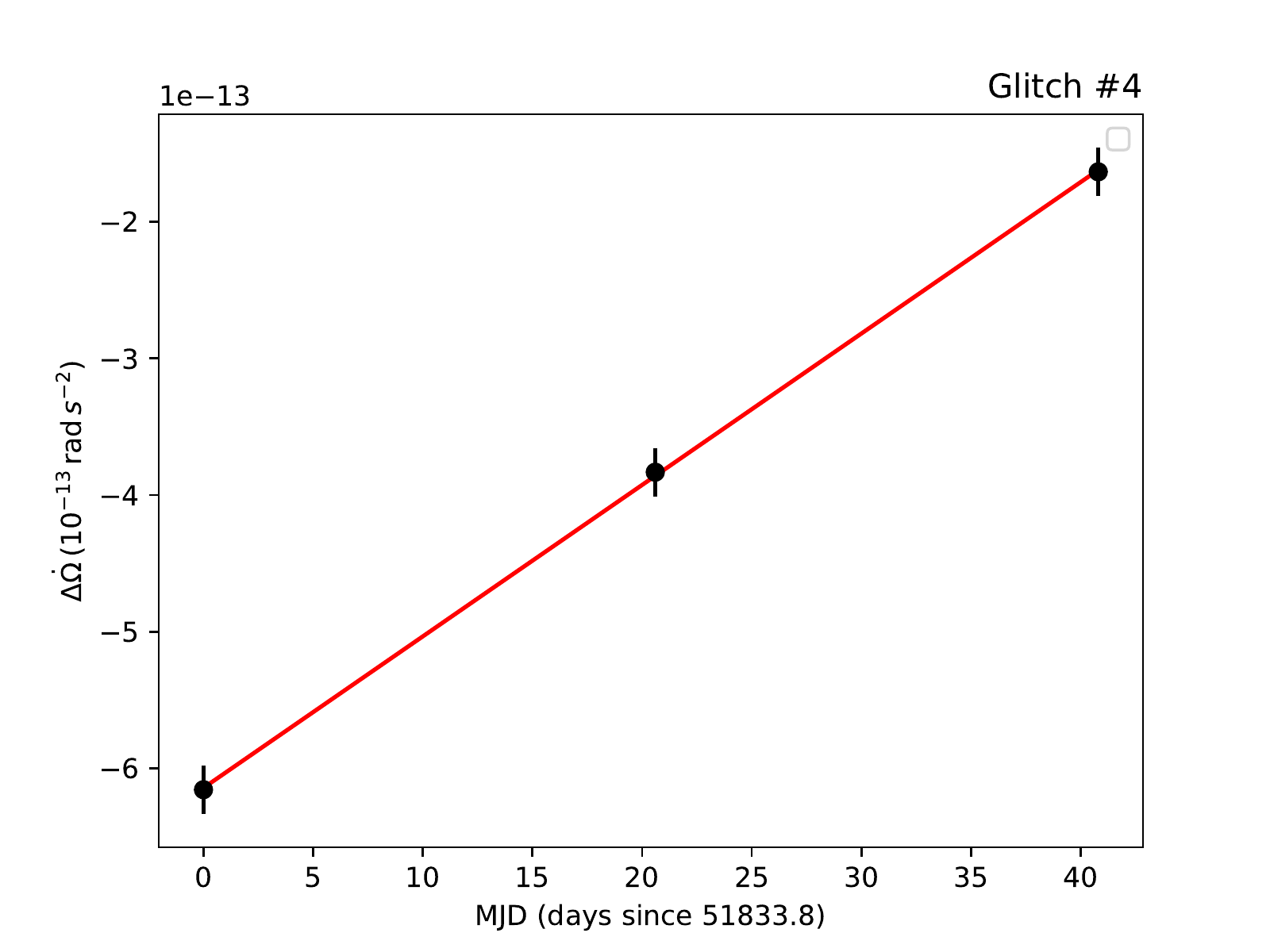}}\
\subfloat[Glitch 5]{\includegraphics[width = 3in,height=5.0cm]{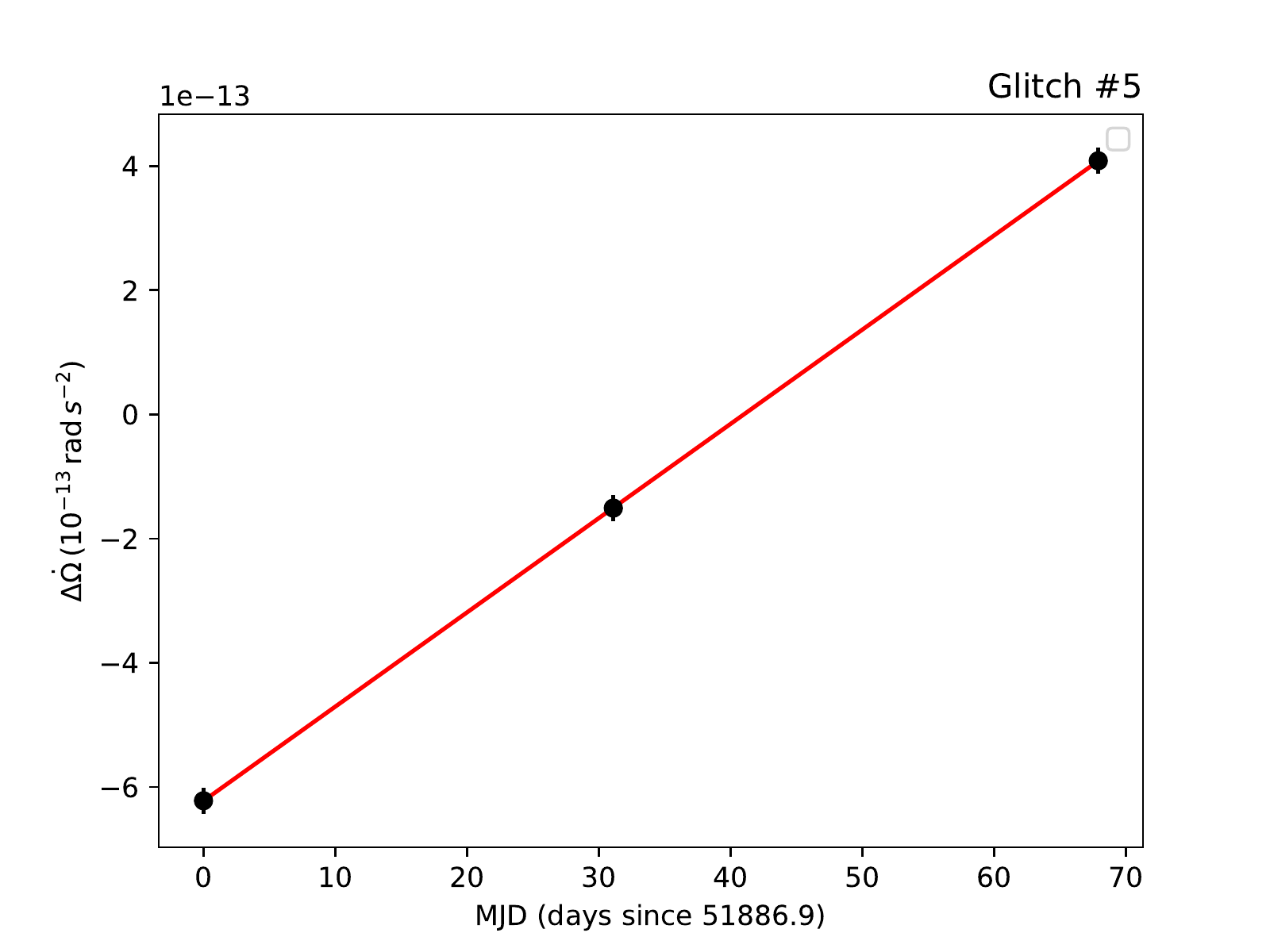}}
\subfloat[Glitch 6]{\includegraphics[width = 3in,height=5.0cm]{G6.pdf}}\
\subfloat[Glitch 8]{\includegraphics[width = 3in,height=5.0cm]{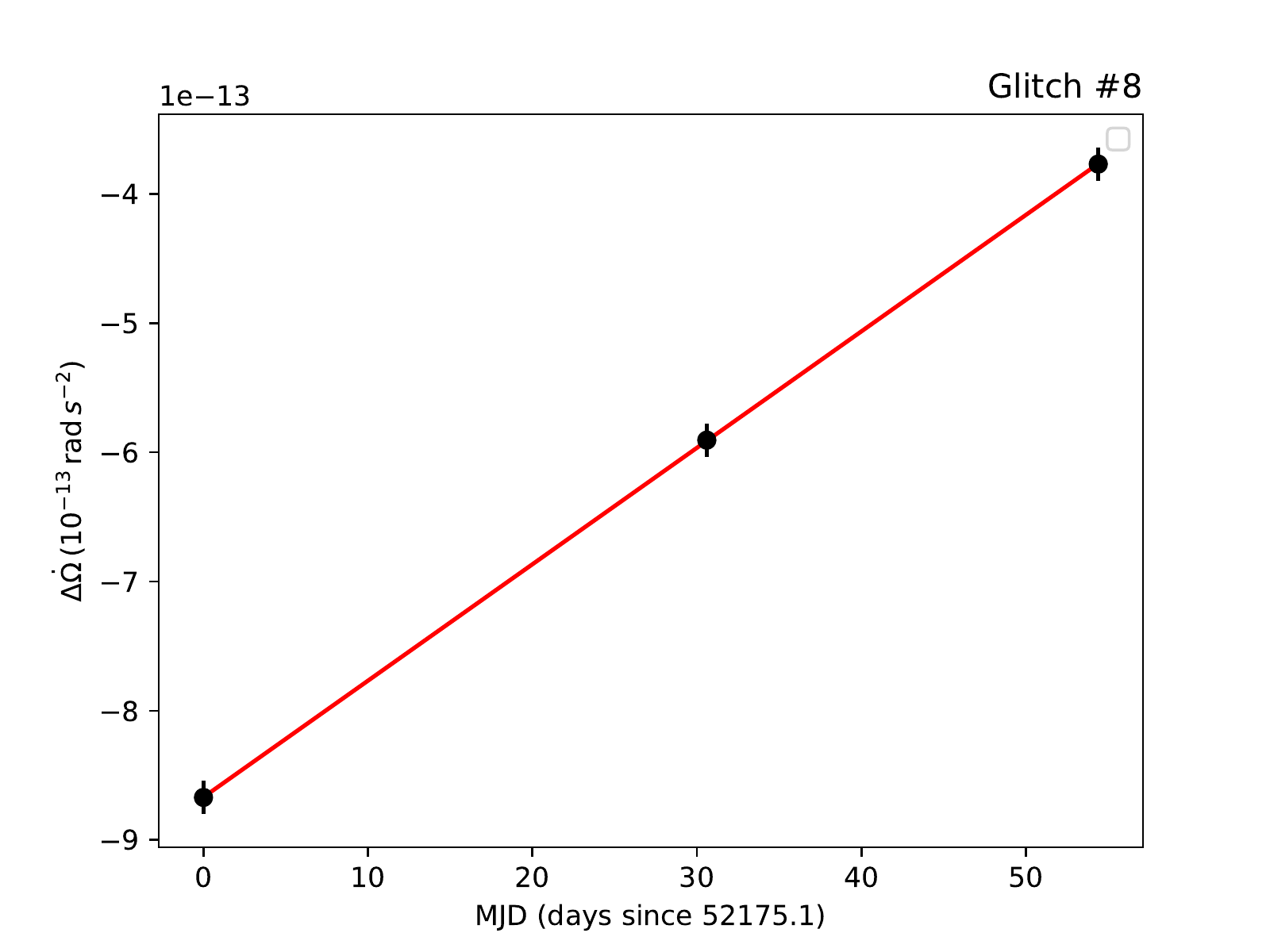}}
\subfloat[Glitch 9]{\includegraphics[width = 3in,height=5.0cm]{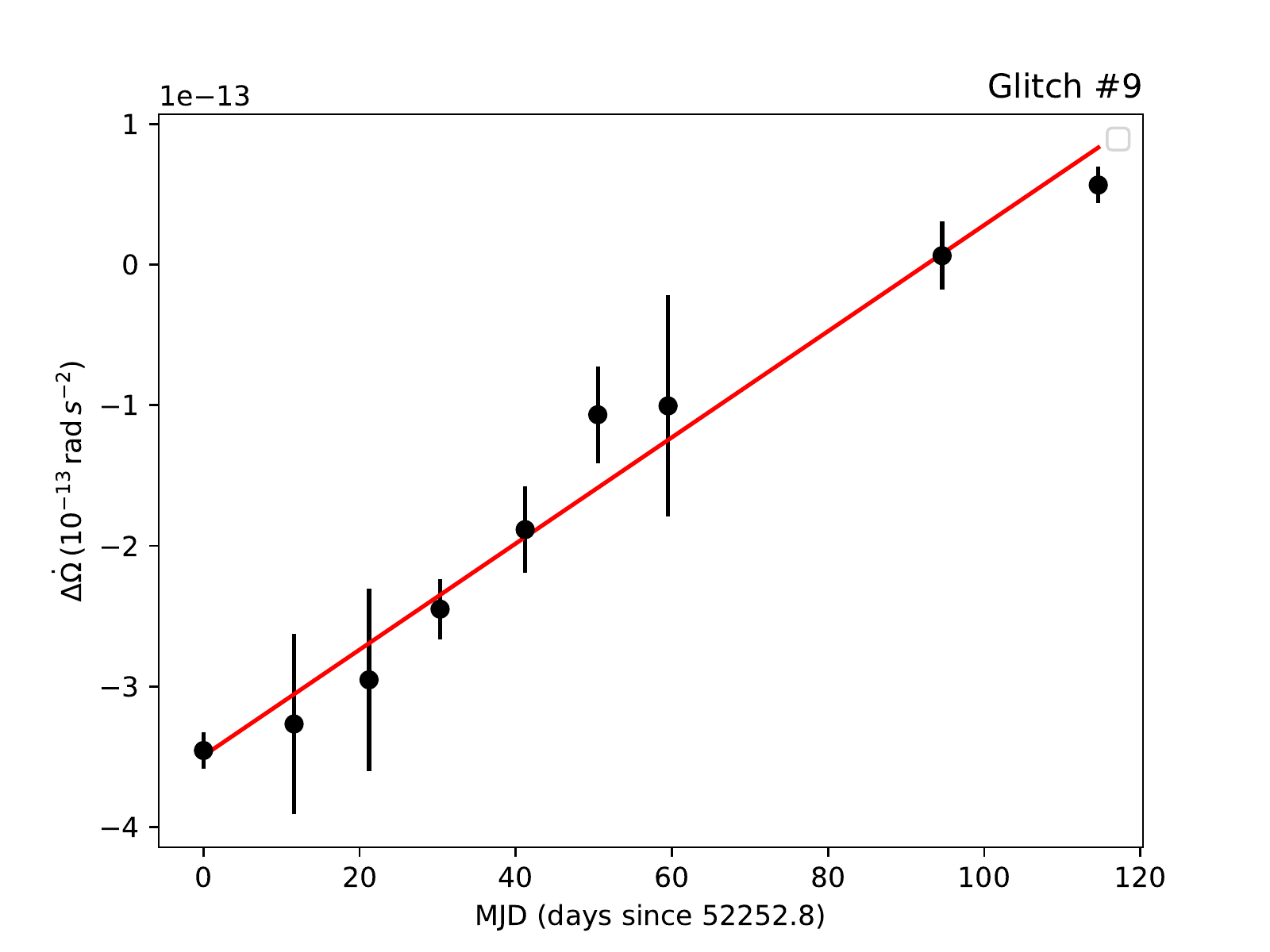}}\

\caption{Model fits to the post-glitch spin-down rate after the 1st, 2nd, 3rd, 4th, 5th, 6th, 8th, and 9th glitches of PSR J0537$-$6910.}
\label{some example}
\end{figure*}

\begin{figure*}
\centering

\subfloat[Glitch 10]{\includegraphics[width = 3in,height=5.0cm]{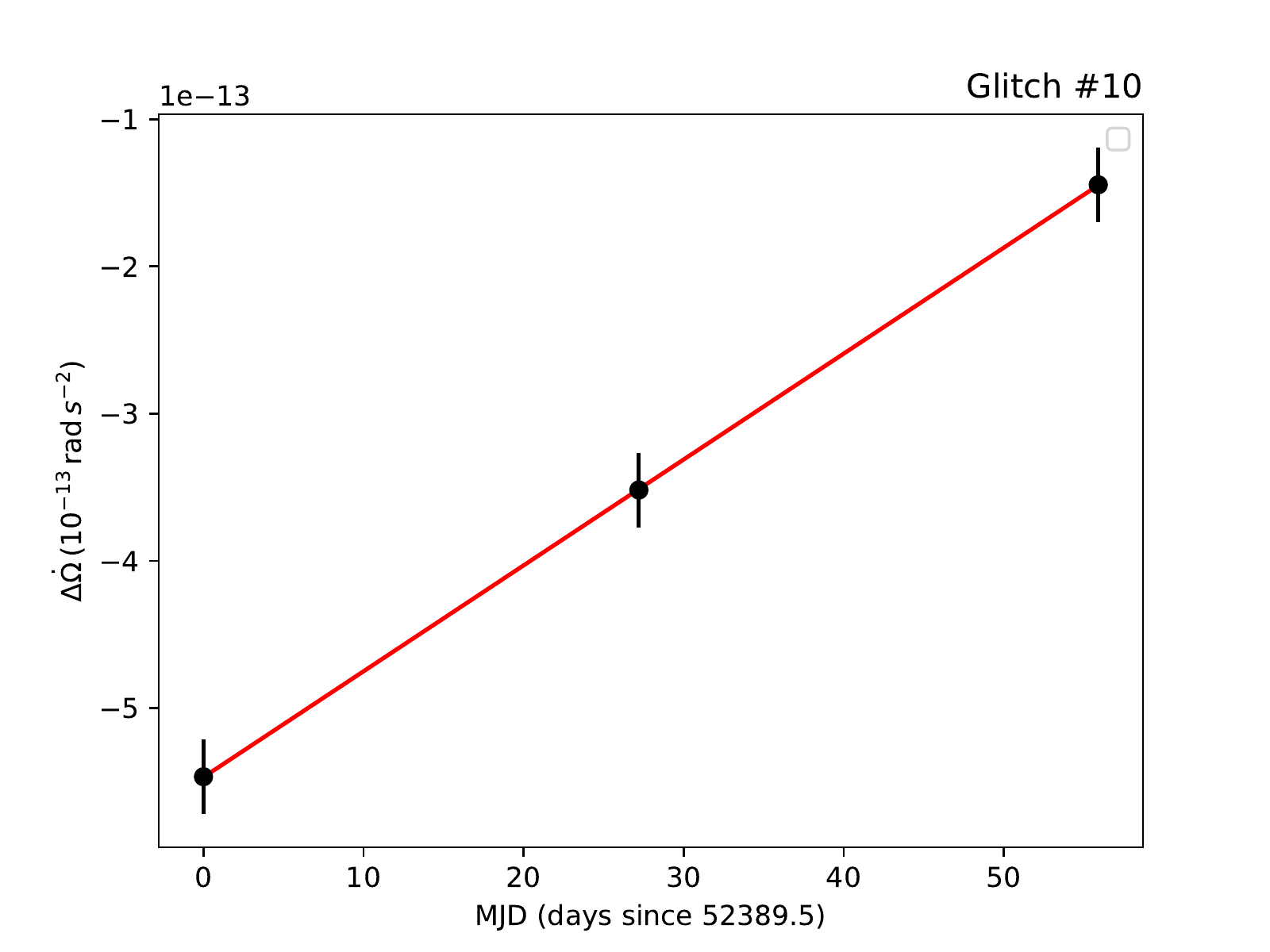}}
\subfloat[Glitch 11]{\includegraphics[width = 3in,height=5.0cm]{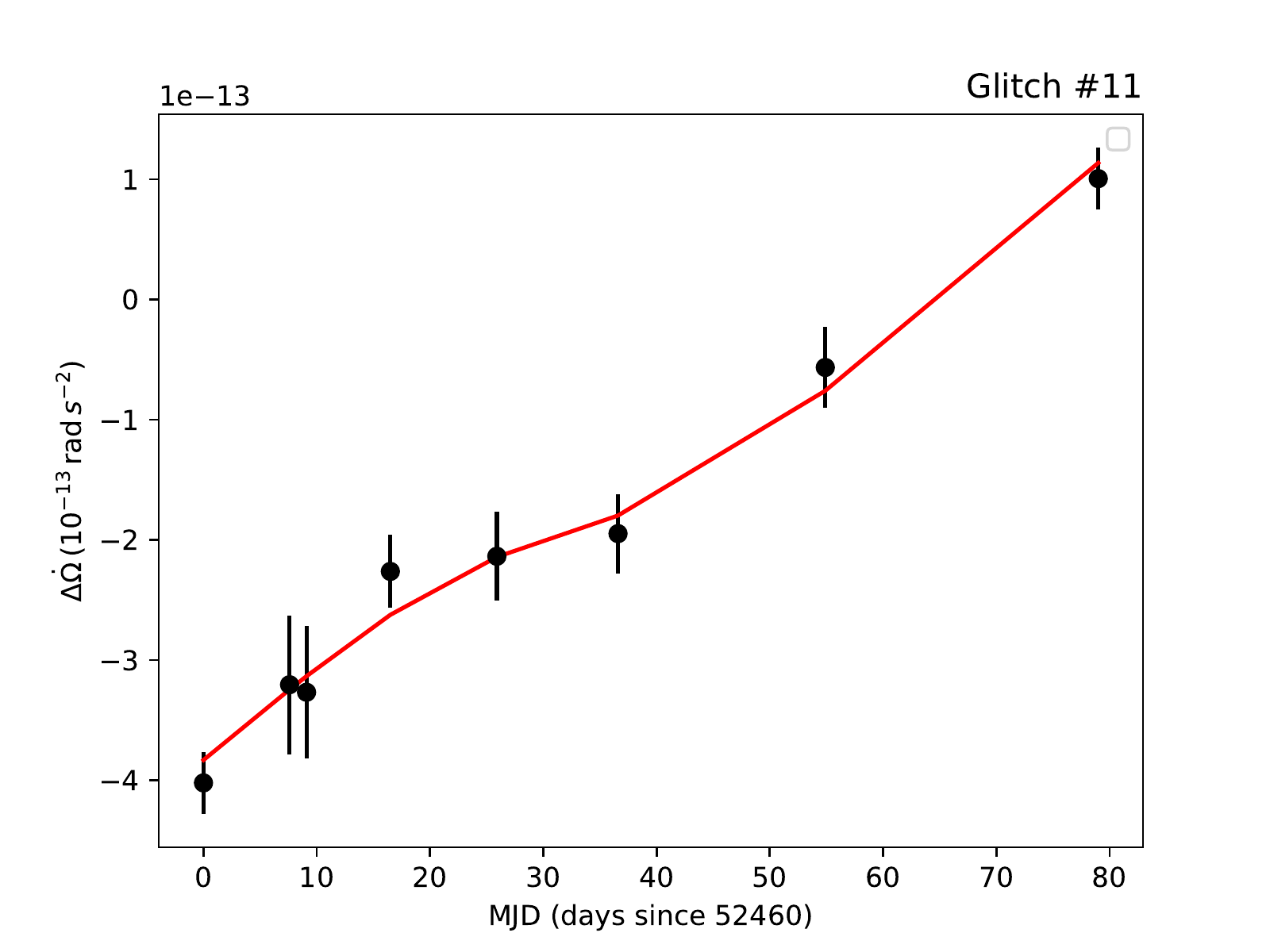}}\
\subfloat[Glitch 12]{\includegraphics[width = 3in,height=5.0cm]{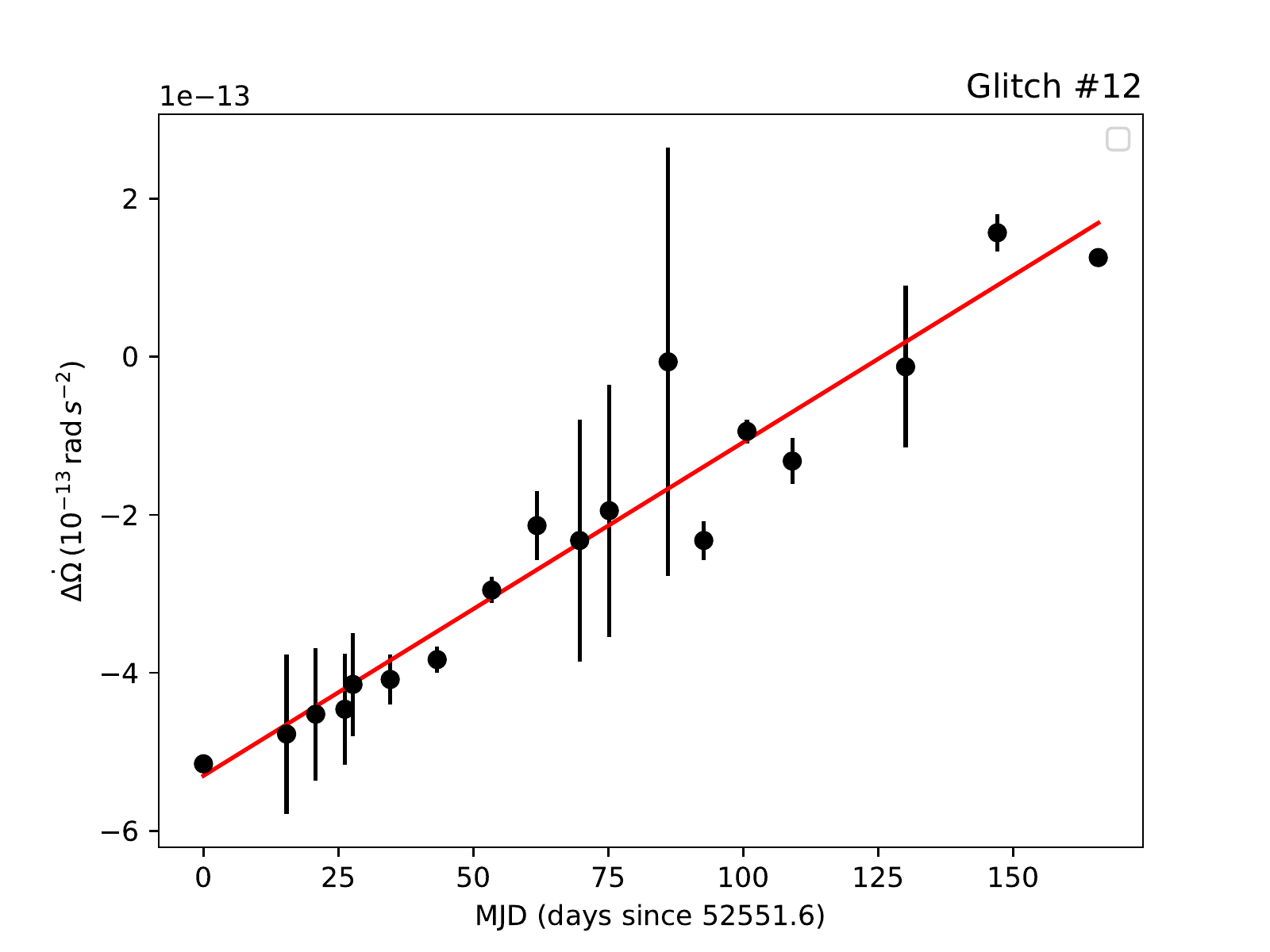}}
\subfloat[Glitch 13]{\includegraphics[width = 3in,height=5.0cm]{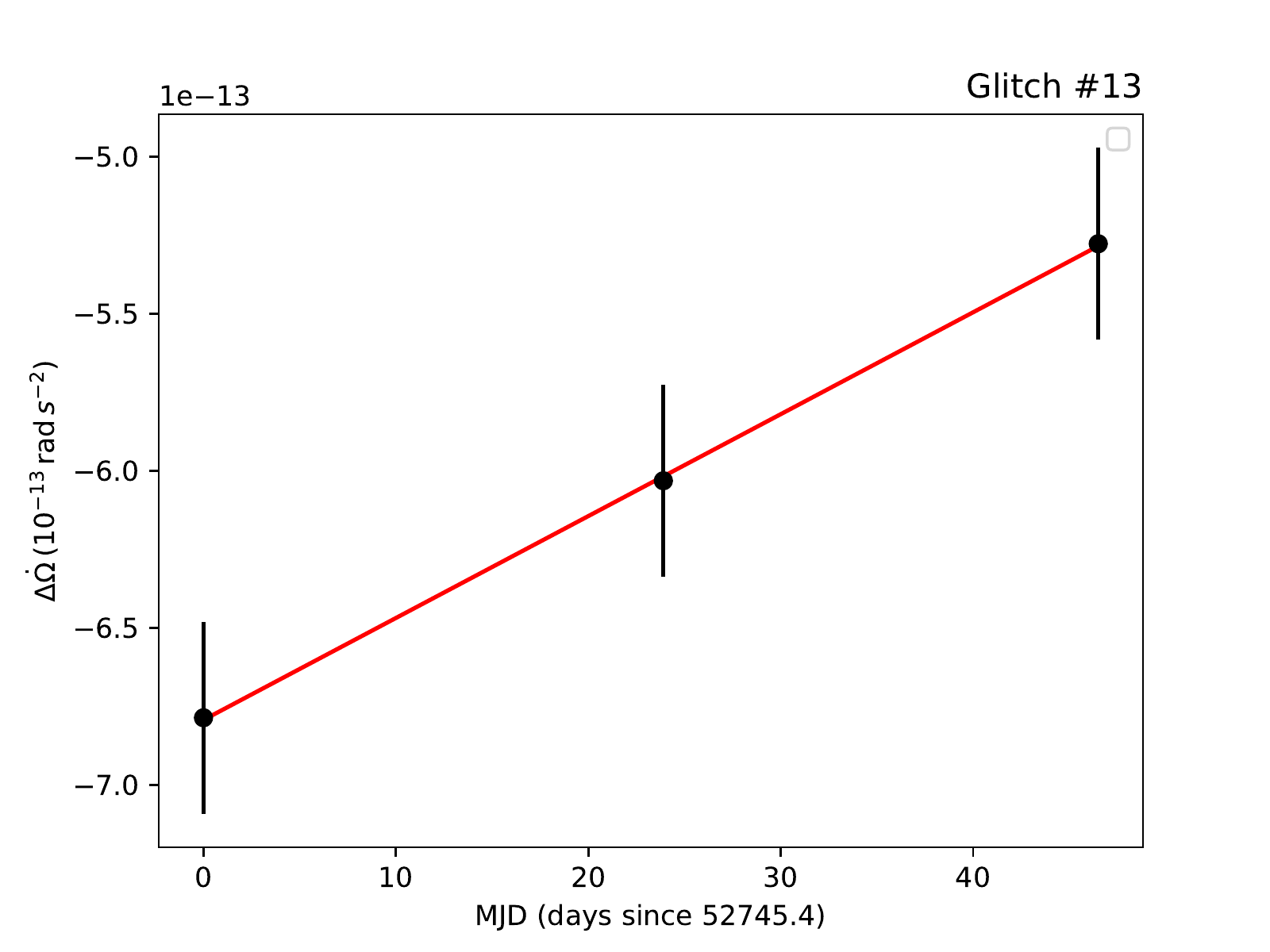}}\
\subfloat[Glitch 14]{\includegraphics[width = 3in,height=5.0cm]{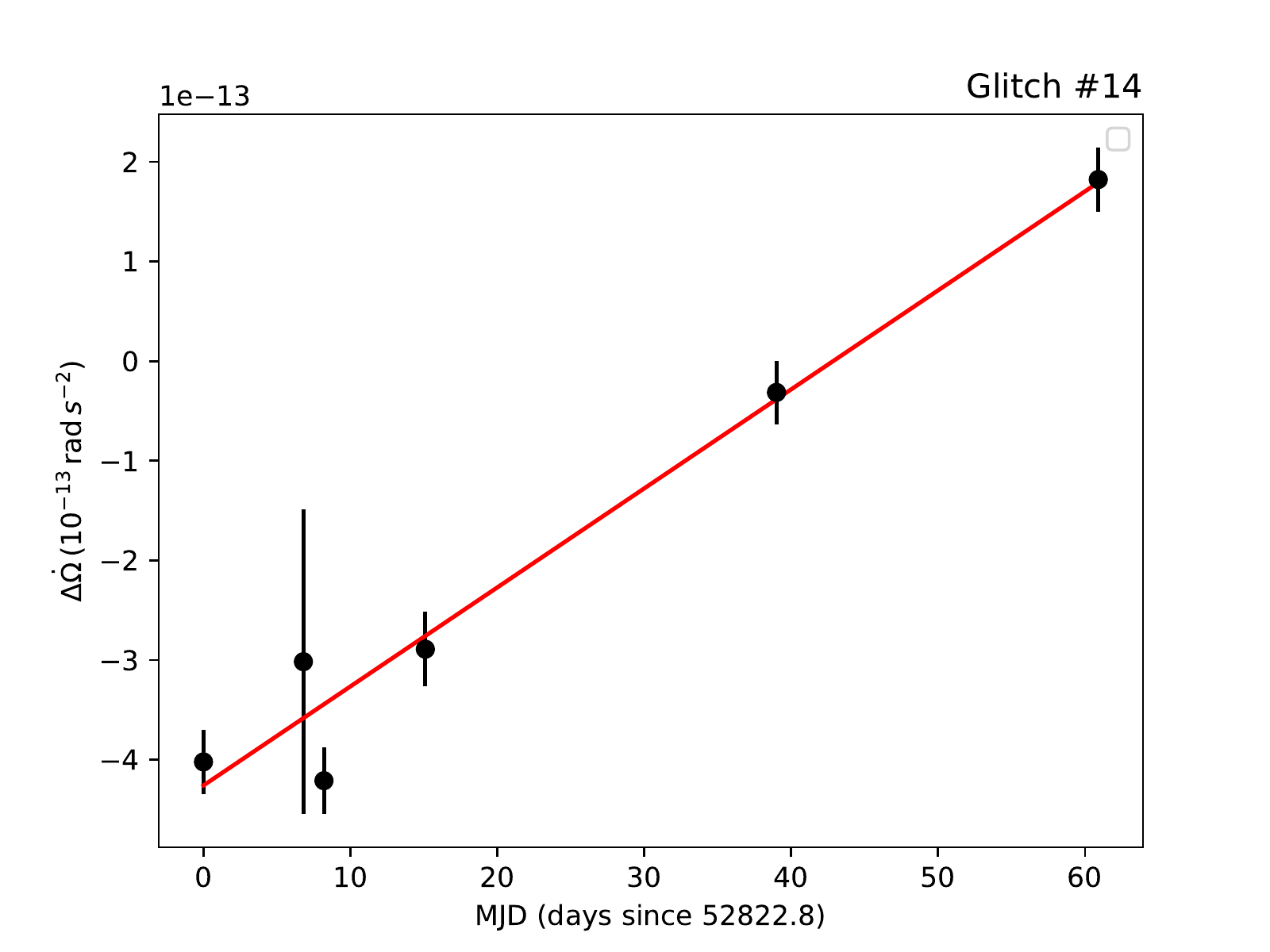}}
\subfloat[Glitch 15]{\includegraphics[width = 3in,height=5.0cm]{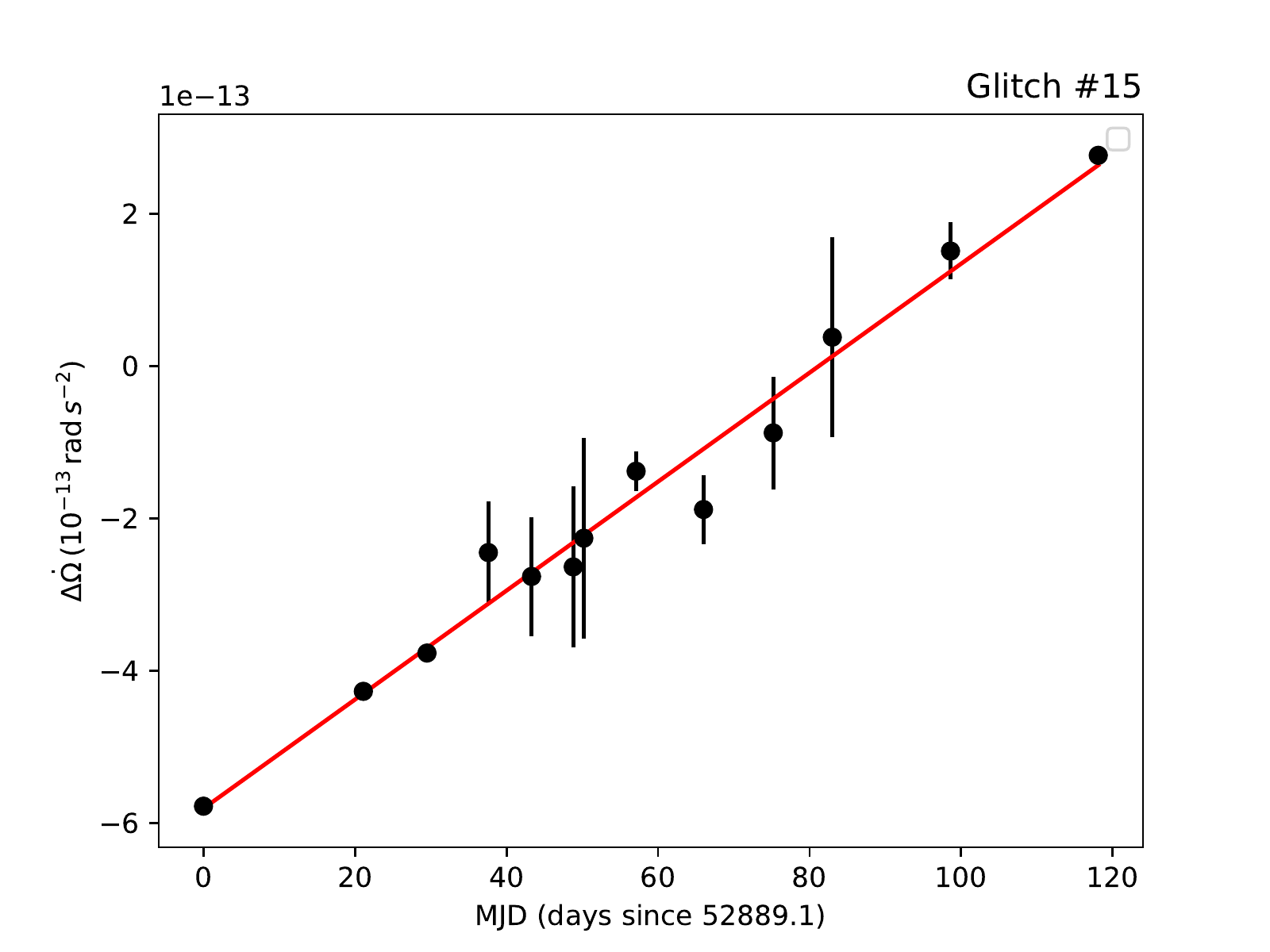}}\
\subfloat[Glitch 16]{\includegraphics[width = 3in,height=5.0cm]{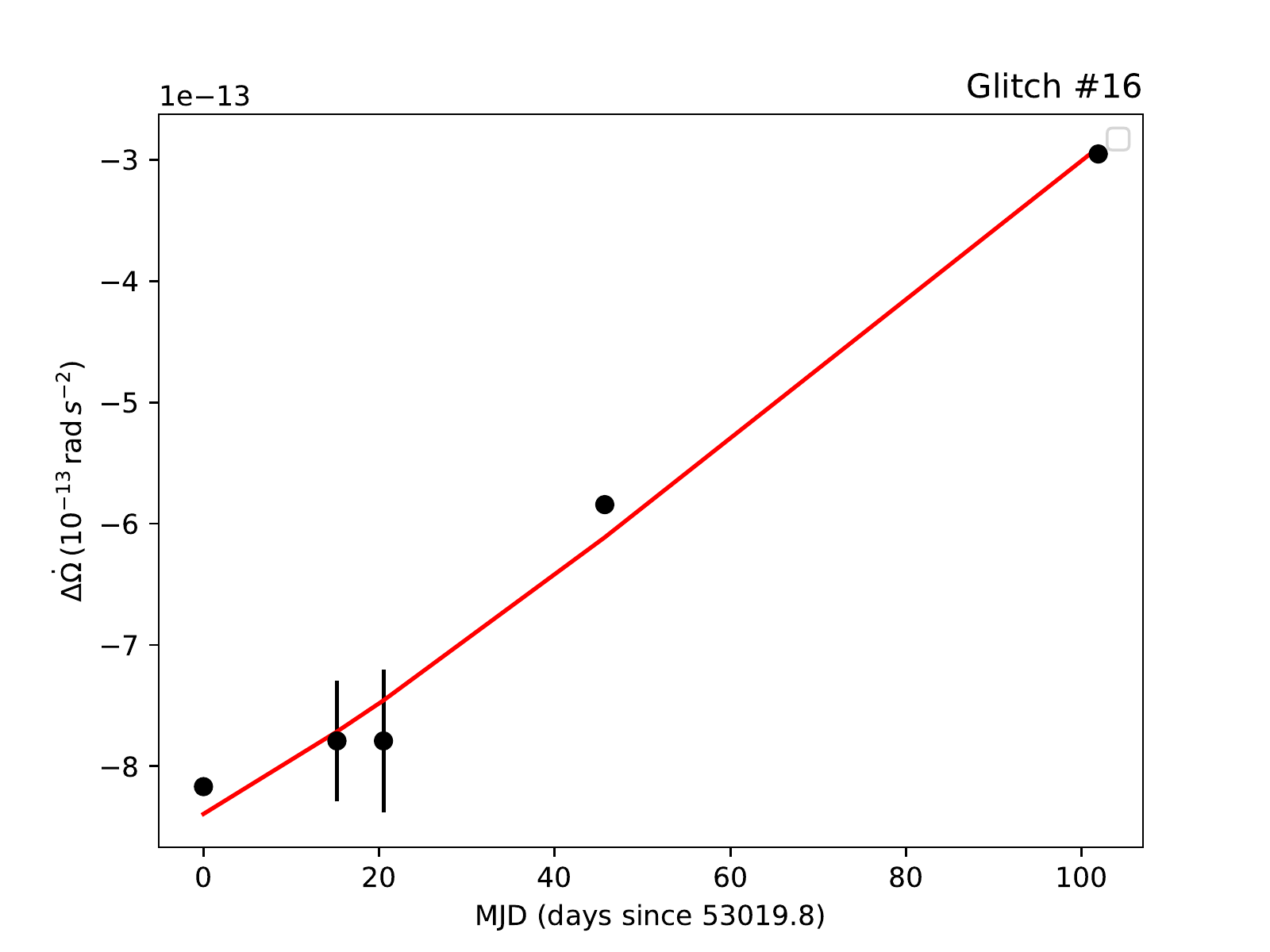}}
\subfloat[Glitch 18]{\includegraphics[width = 3in,height=5.0cm]{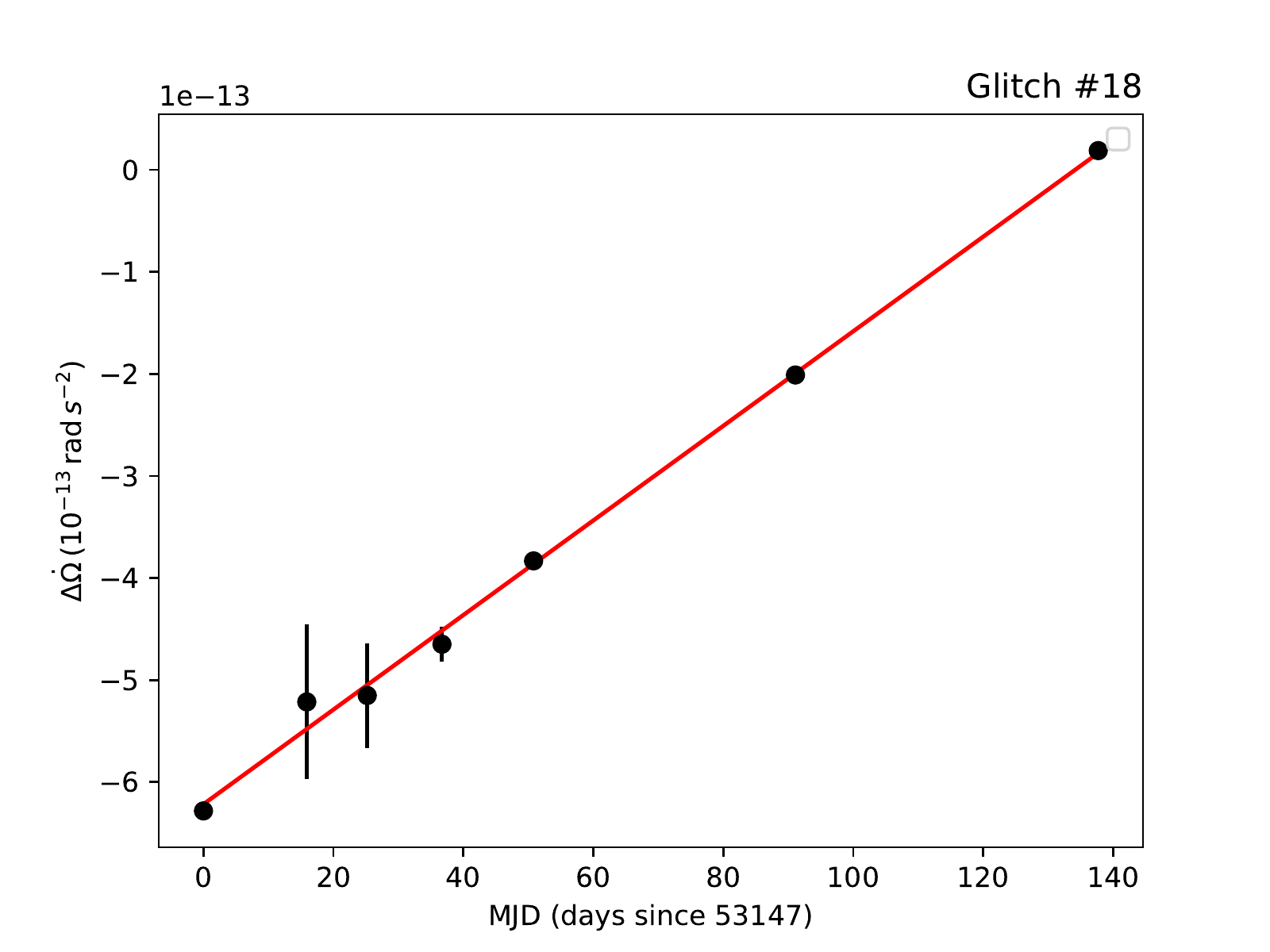}}\

\caption{Model fits to the post-glitch spin-down rate after the 10th, 11th, 12th, 13th, 14th, 15th, 16th, and 18th glitches of PSR J0537$-$6910.}
\label{some example}
\end{figure*}

\begin{figure*}
\centering

\subfloat[Glitch 20]{\includegraphics[width = 3in,height=5.0cm]{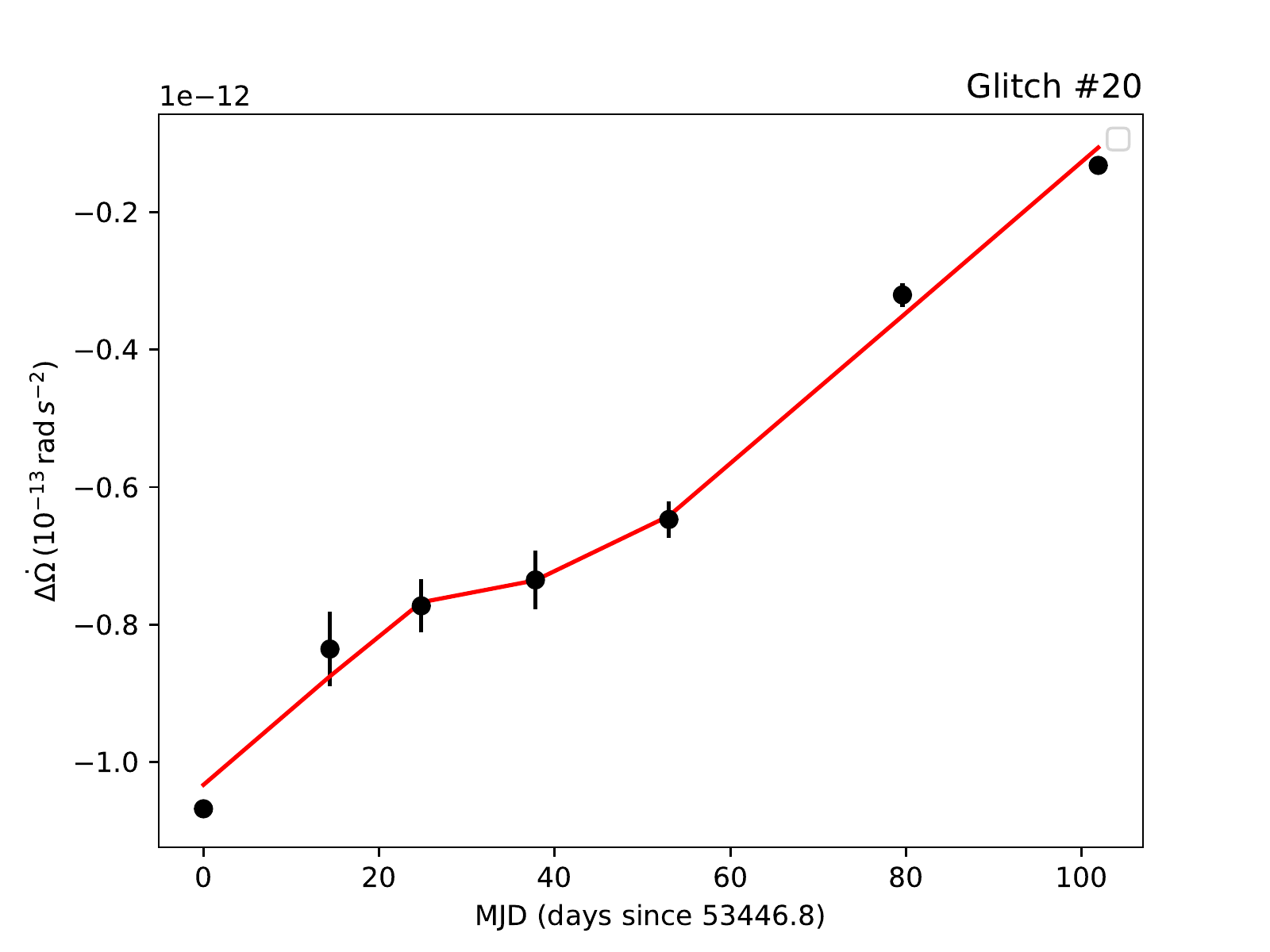}}
\subfloat[Glitch 21]{\includegraphics[width = 3in,height=5.0cm]{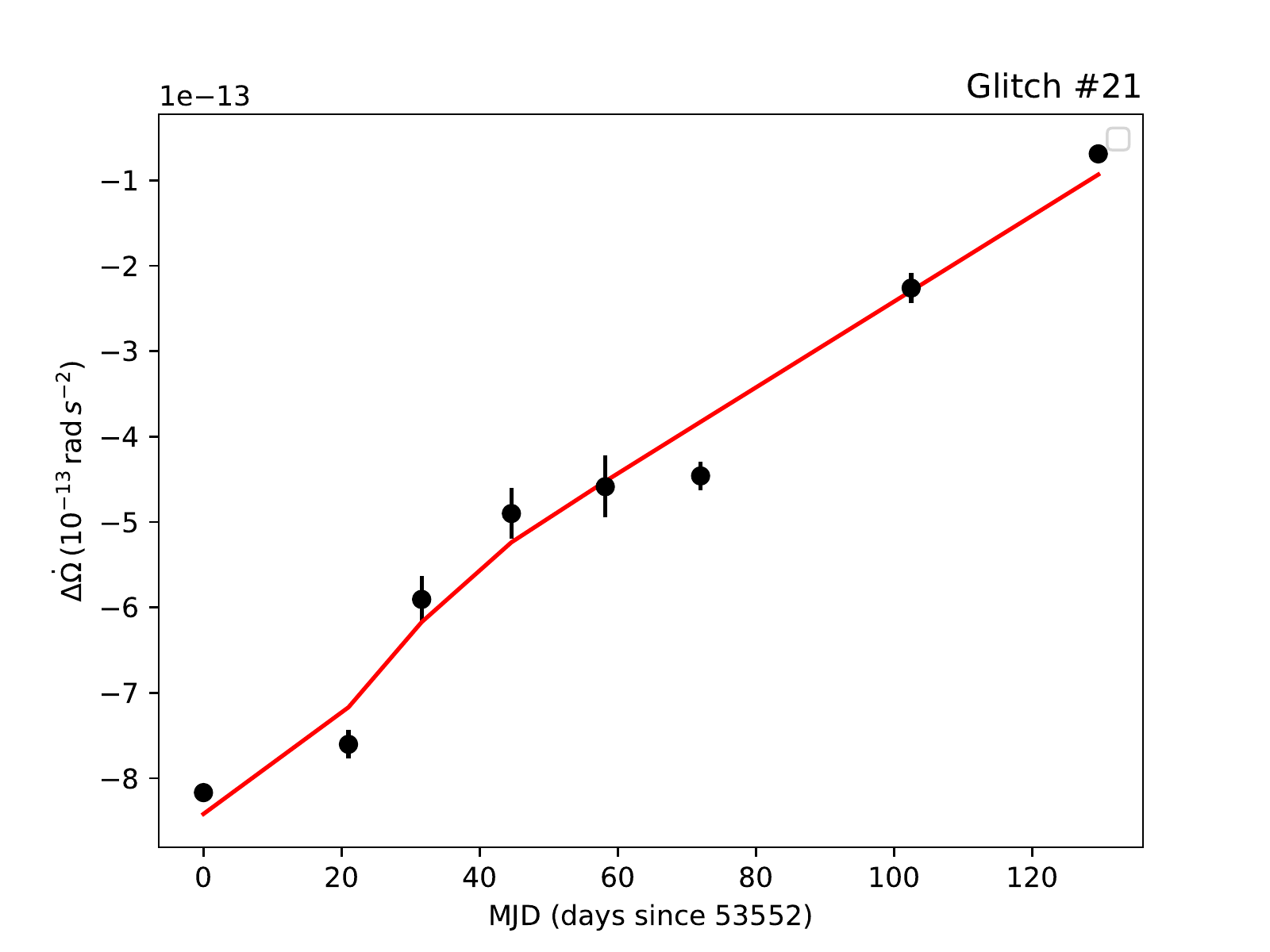}}\
\subfloat[Glitch 22]{\includegraphics[width = 3in,height=5.0cm]{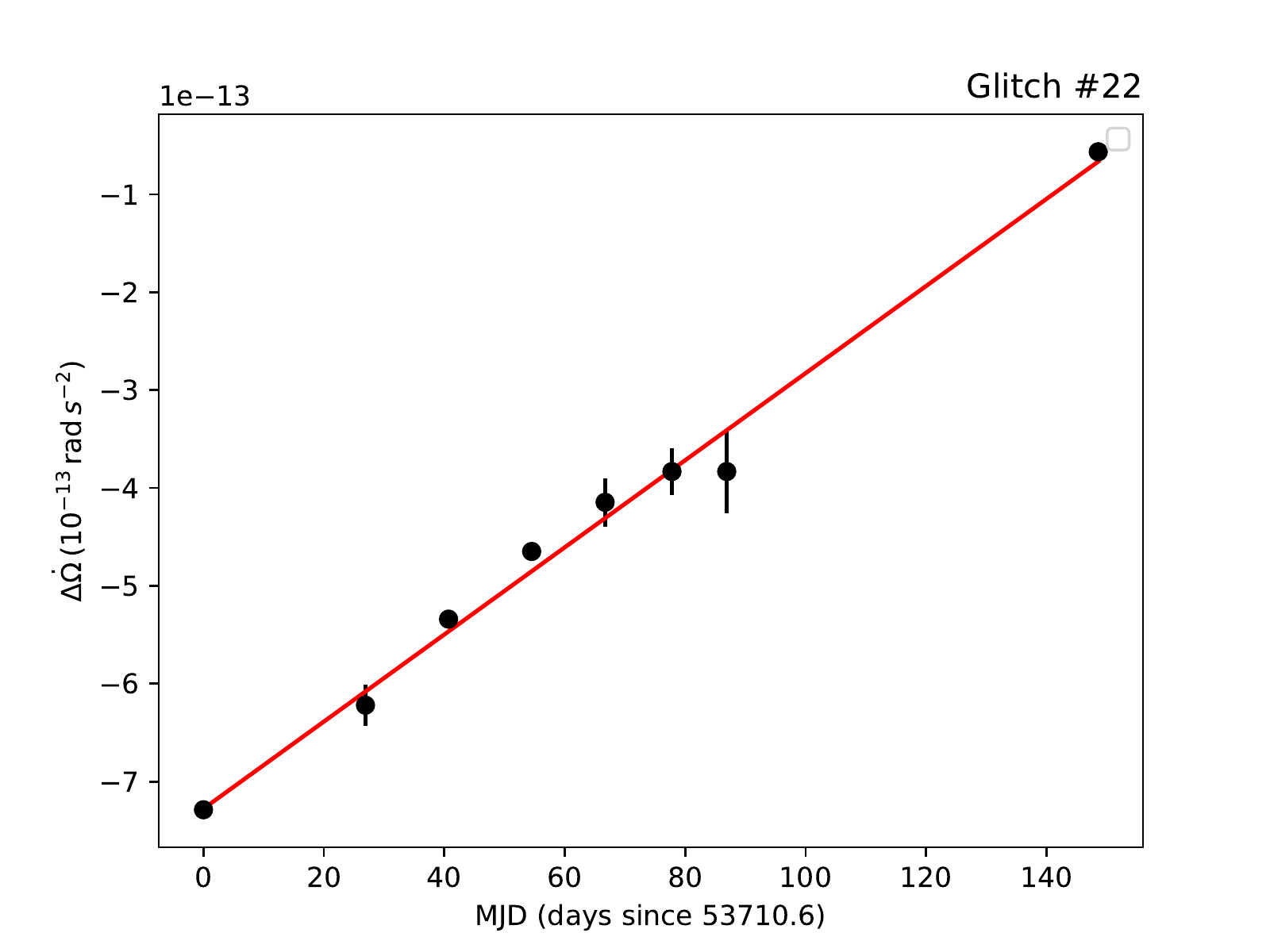}}
\subfloat[Glitch 23]{\includegraphics[width = 3in,height=5.0cm]{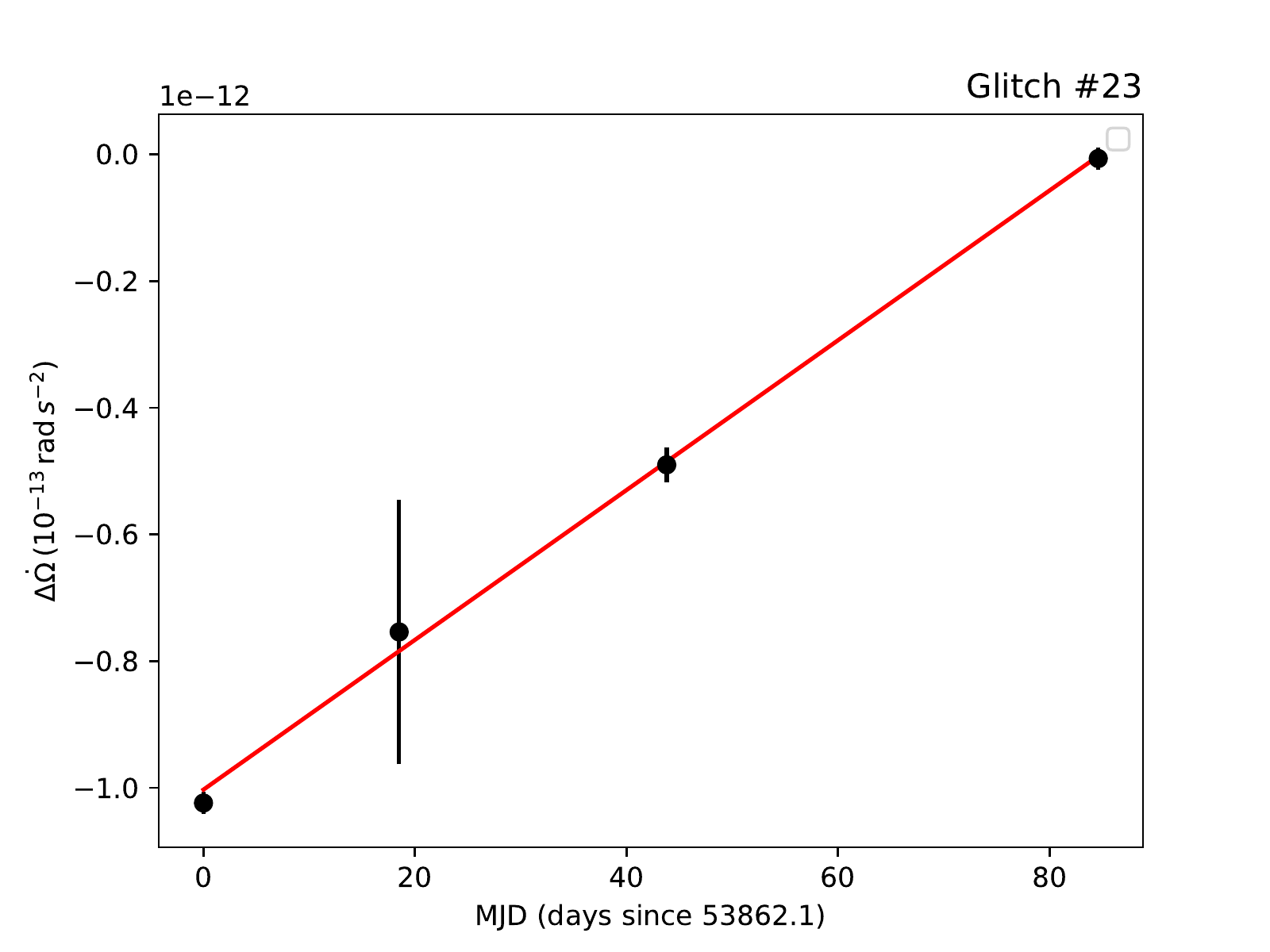}}\
\subfloat[Glitch 24]{\includegraphics[width = 3in,height=5.0cm]{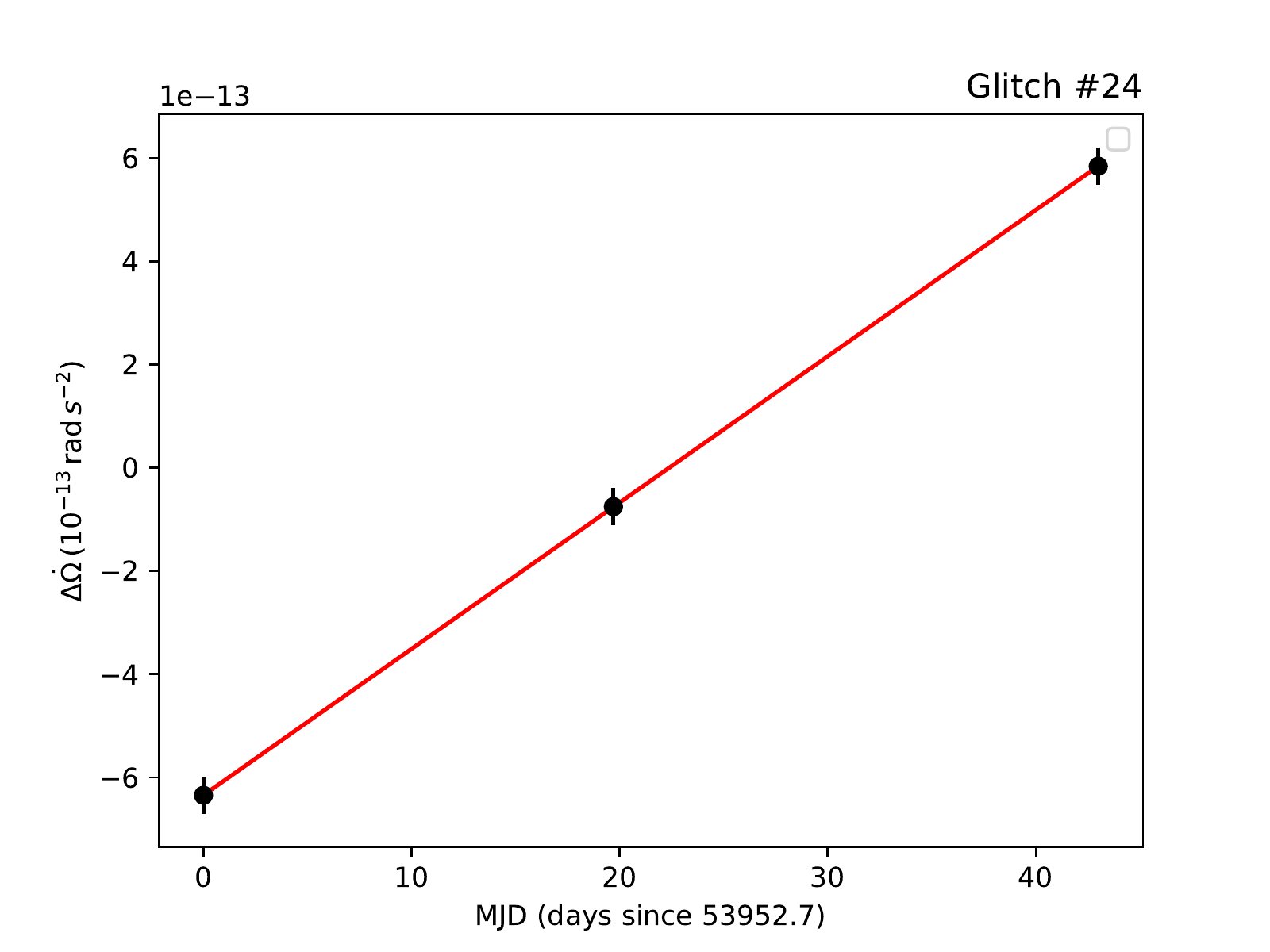}}
\subfloat[Glitch 25]{\includegraphics[width = 3in,height=5.0cm]{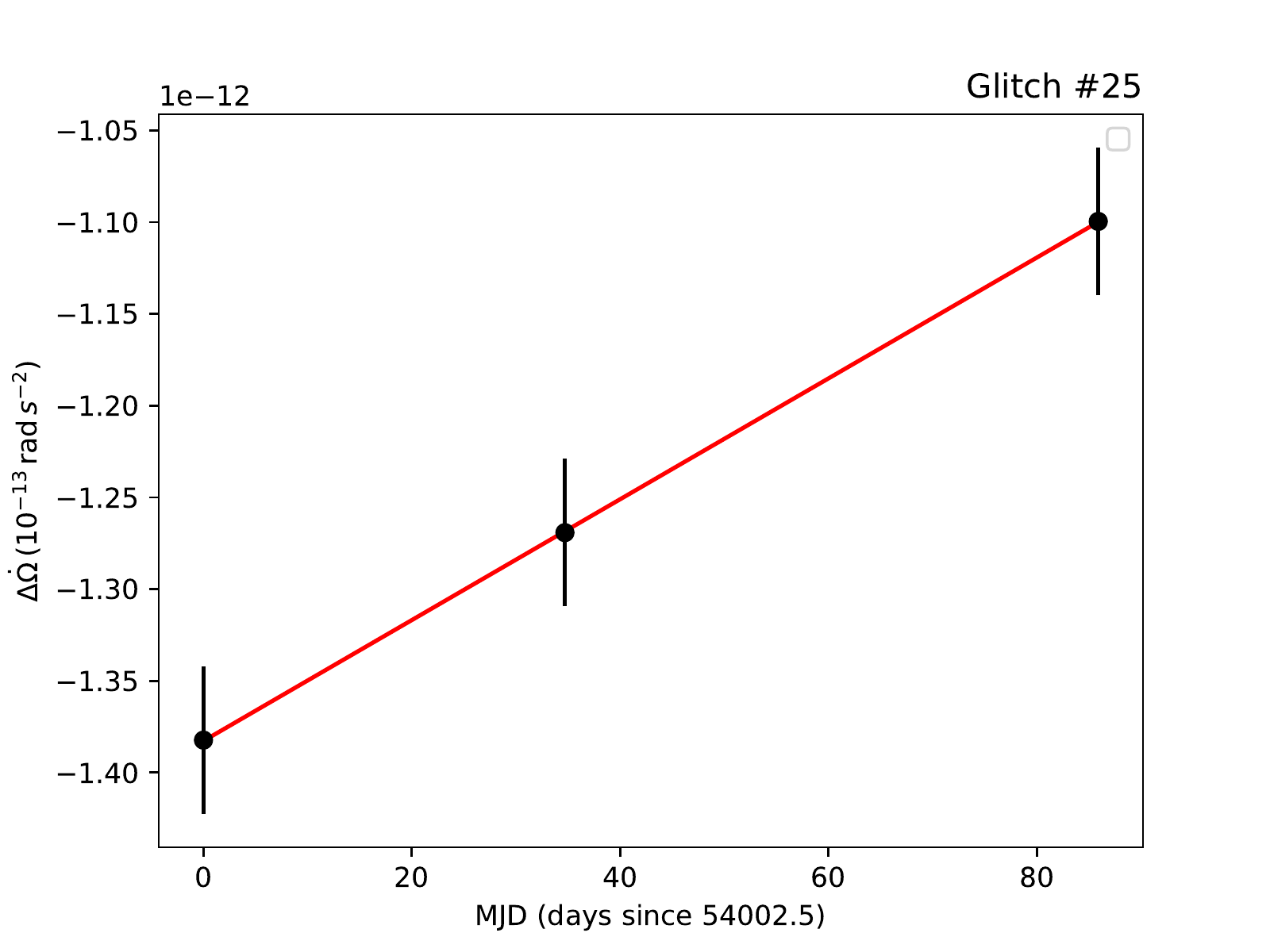}}\
\subfloat[Glitch 26]{\includegraphics[width = 3in,height=5.0cm]{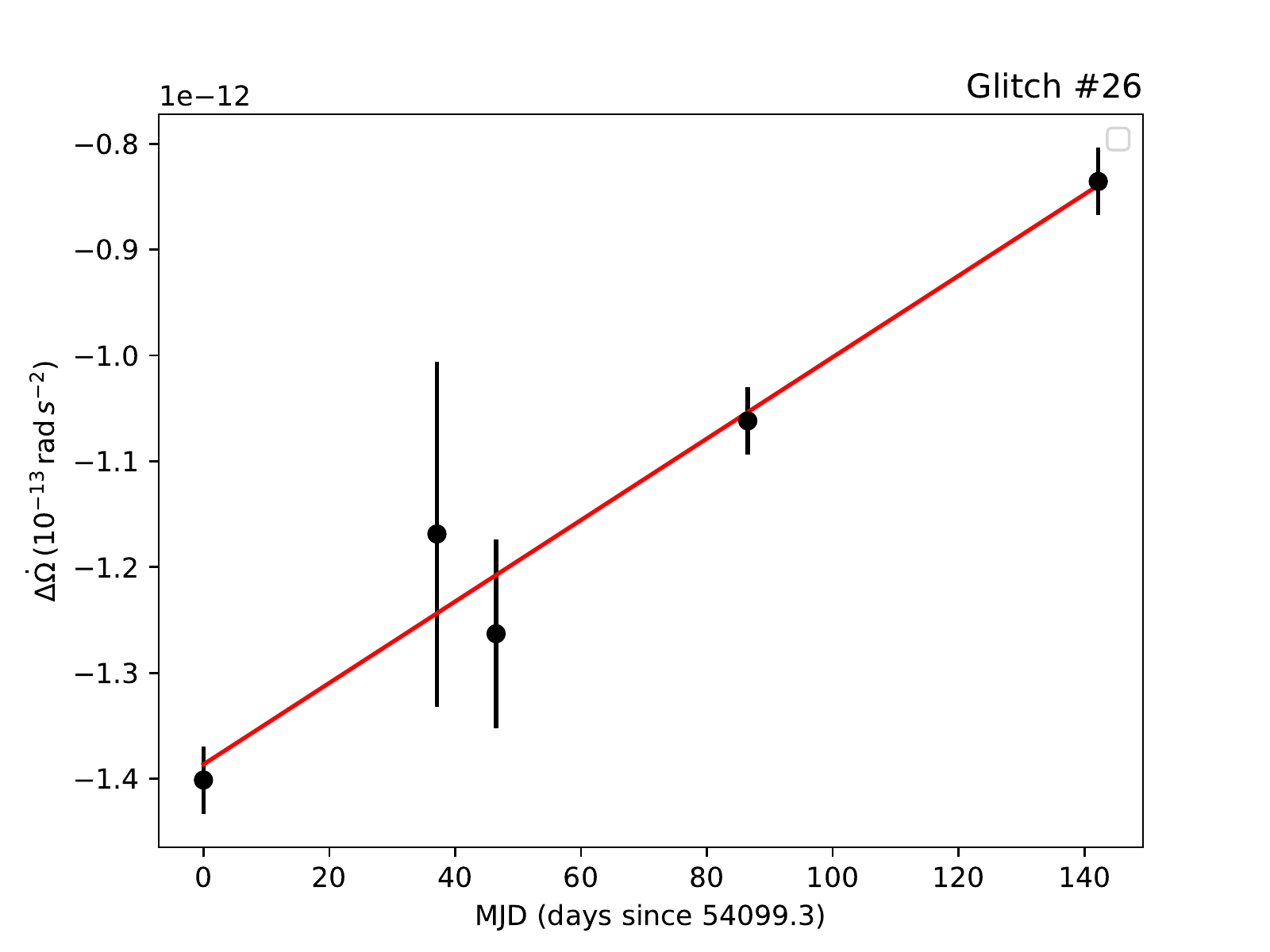}}
\subfloat[Glitch 28]{\includegraphics[width = 3in,height=5.0cm]{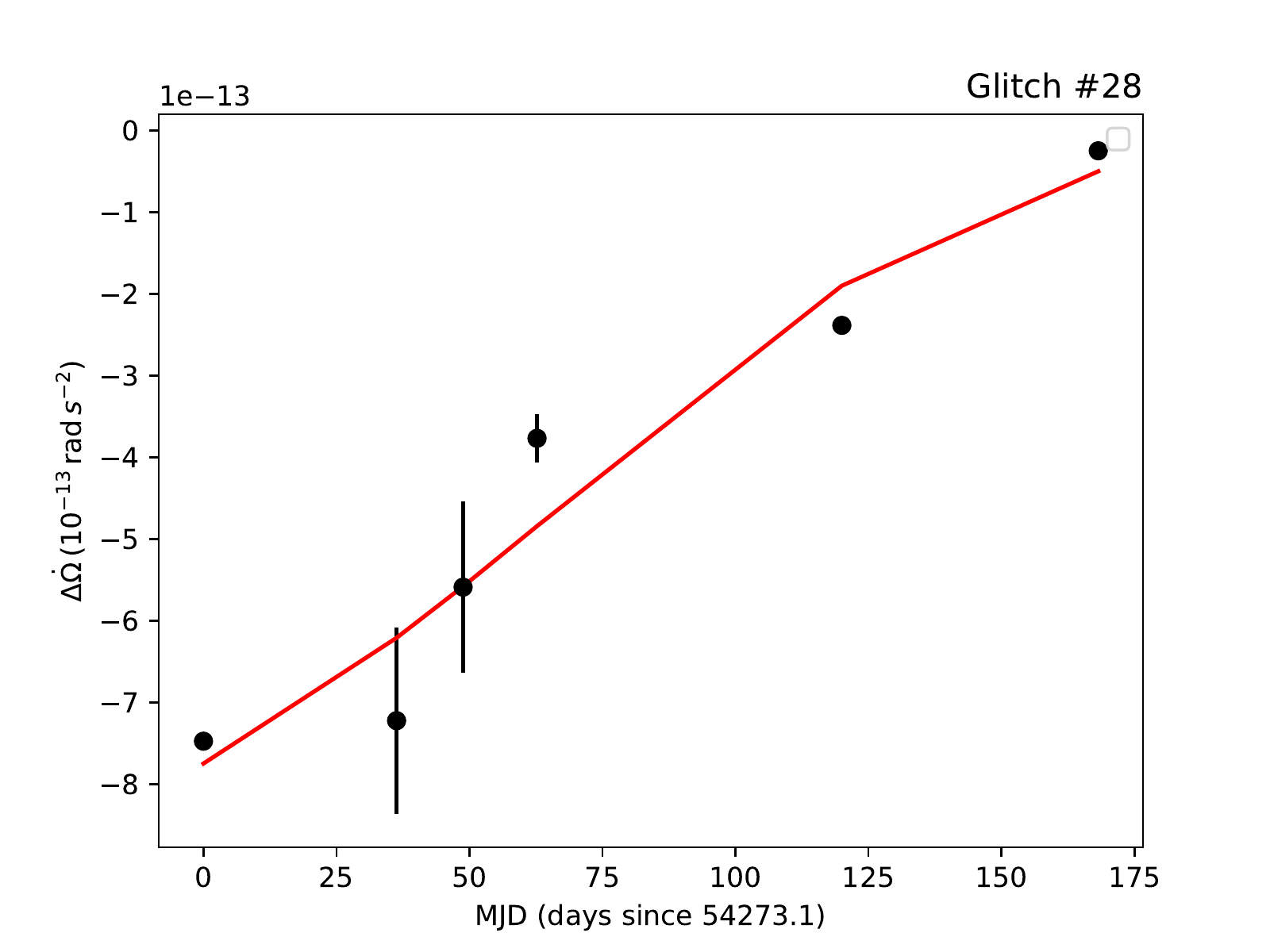}}\

\caption{Model fits to the post-glitch spin-down rate after the 20th, 21st, 22nd, 23th, 24th, 25th, 26th, and 28th glitches of PSR J0537$-$6910.}
\label{some example}
\end{figure*}

\begin{figure*}
\centering

\subfloat[Glitch 29]{\includegraphics[width = 3in,height=5.0cm]{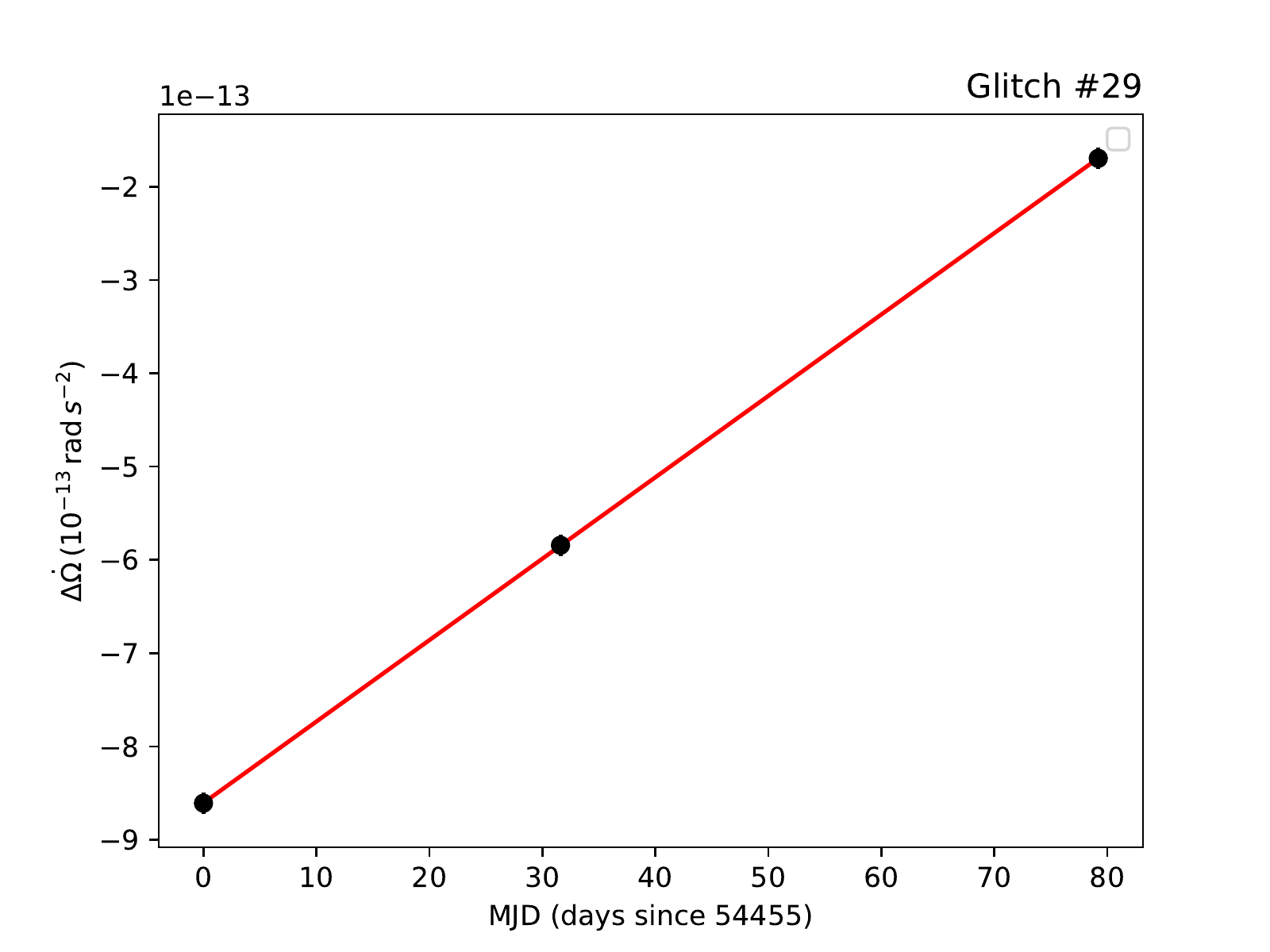}}
\subfloat[Glitch 33]{\includegraphics[width = 3in,height=5.0cm]{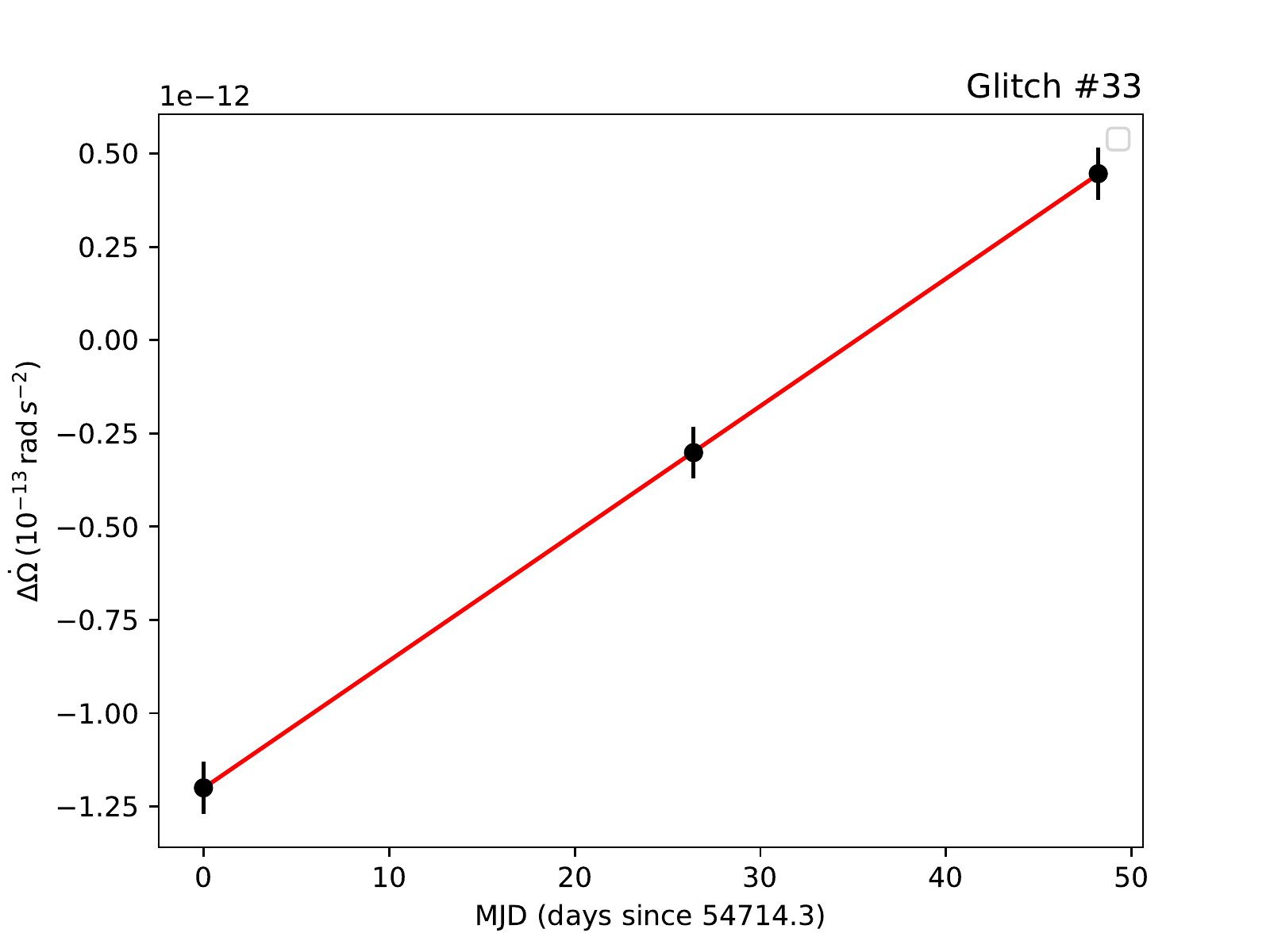}}\
\subfloat[Glitch 34]{\includegraphics[width = 3in,height=5.0cm]{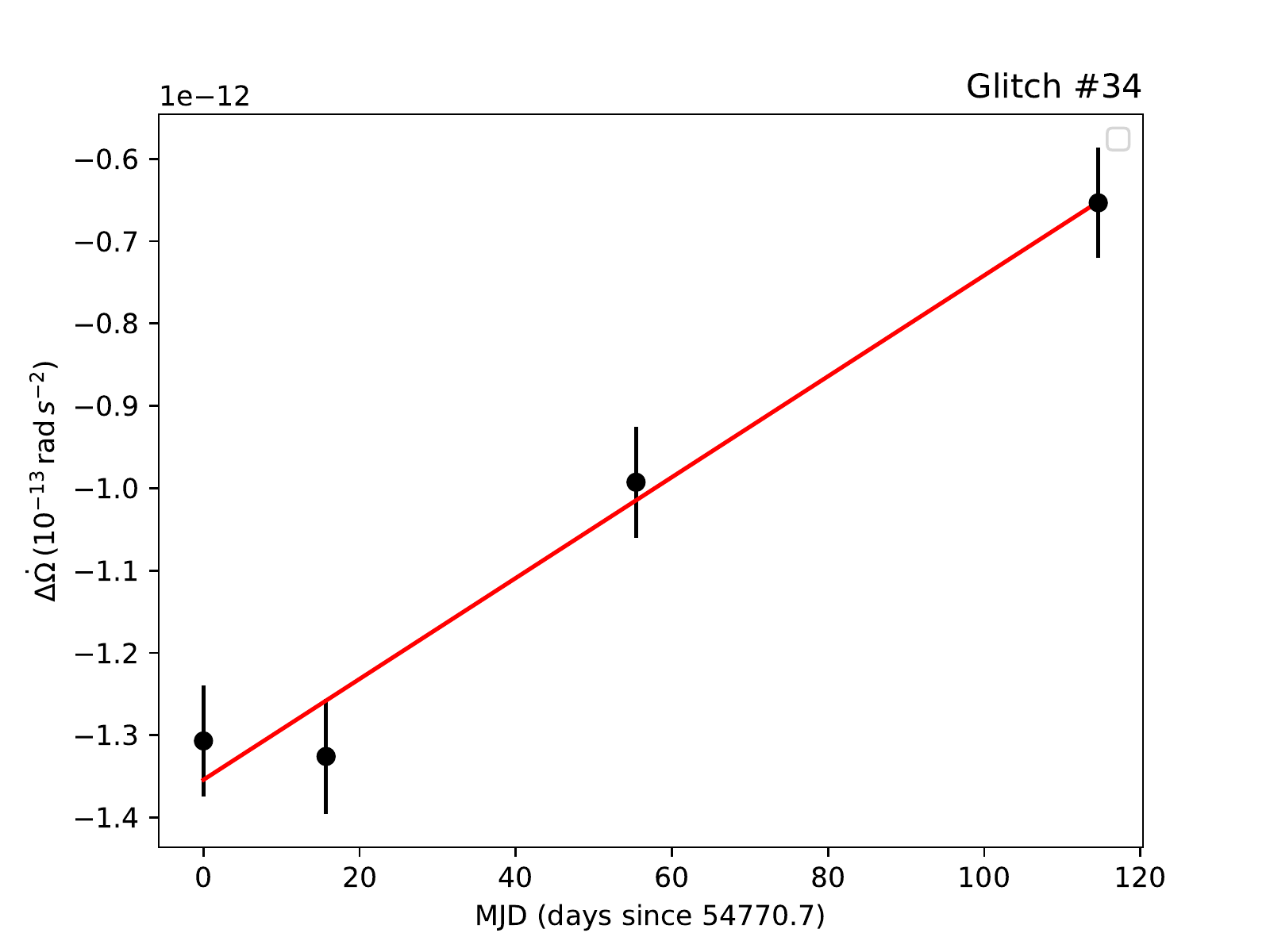}}
\subfloat[Glitch 35]{\includegraphics[width = 3in,height=5.0cm]{G35.pdf}}\
\subfloat[Glitch 36]{\includegraphics[width = 3in,height=5.0cm]{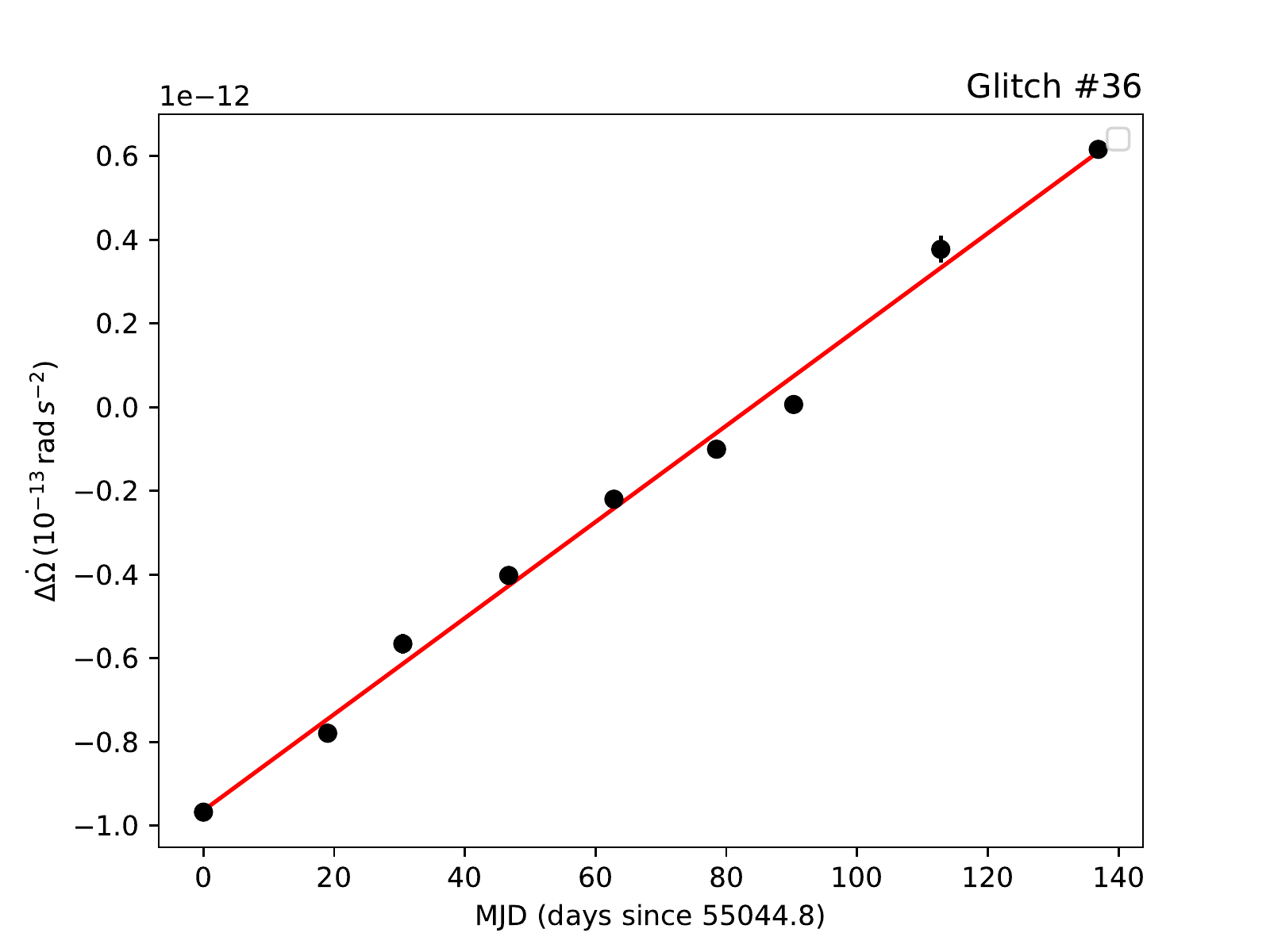}}
\subfloat[Glitch 37]{\includegraphics[width = 3in,height=5.0cm]{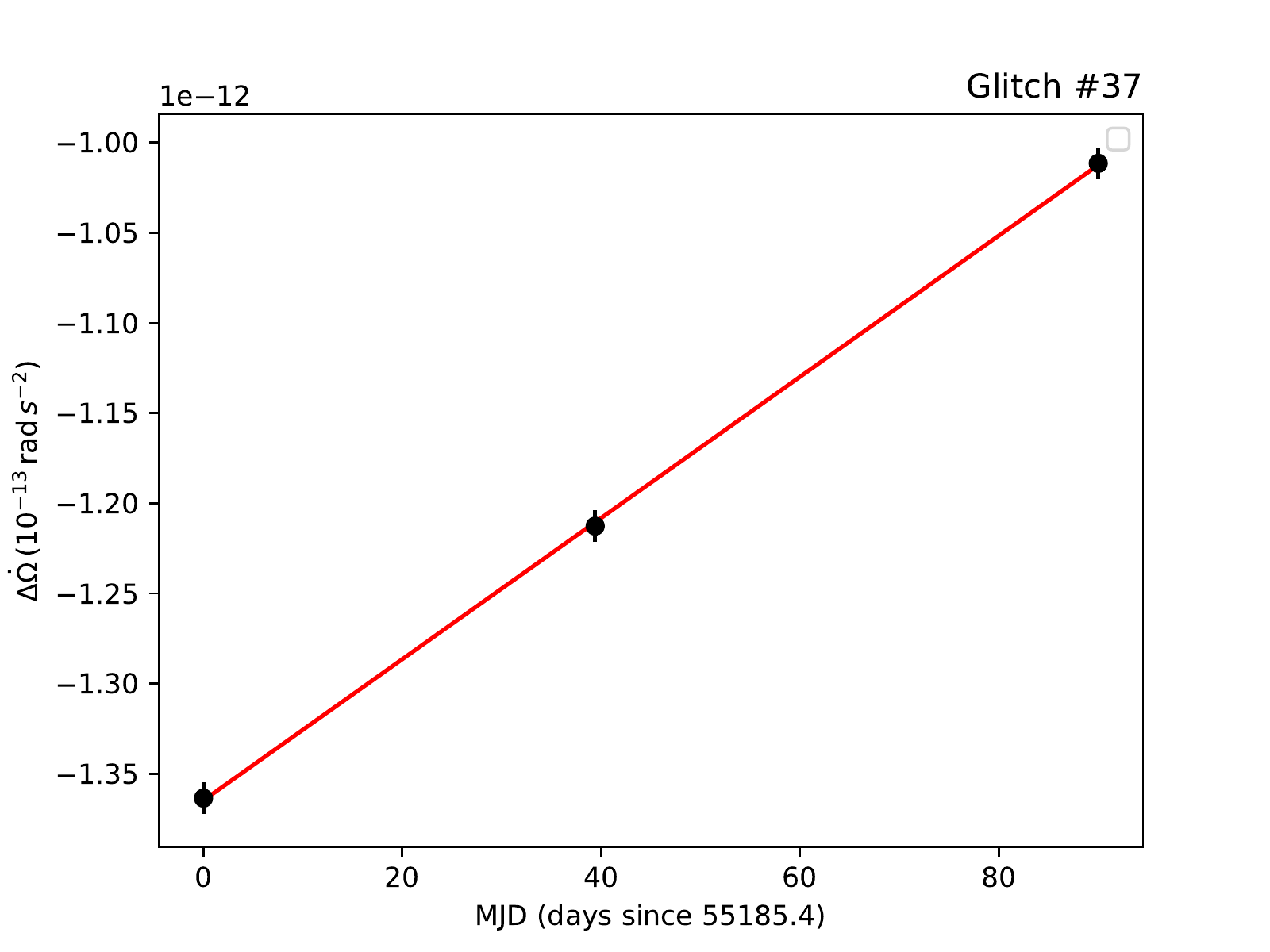}}\
\subfloat[Glitch 38]{\includegraphics[width = 3in,height=5.0cm]{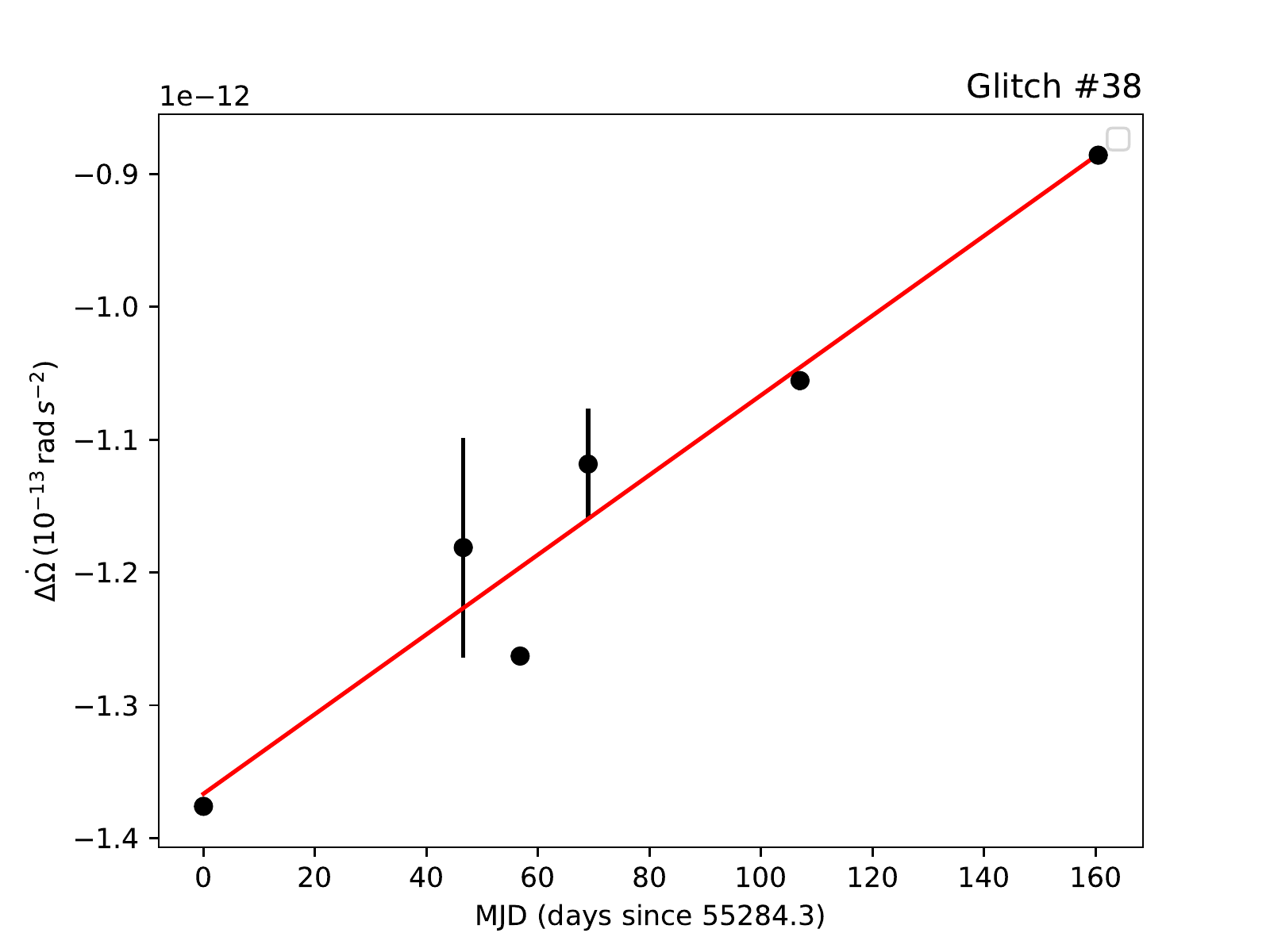}}
\subfloat[Glitch 39]{\includegraphics[width = 3in,height=5.0cm]{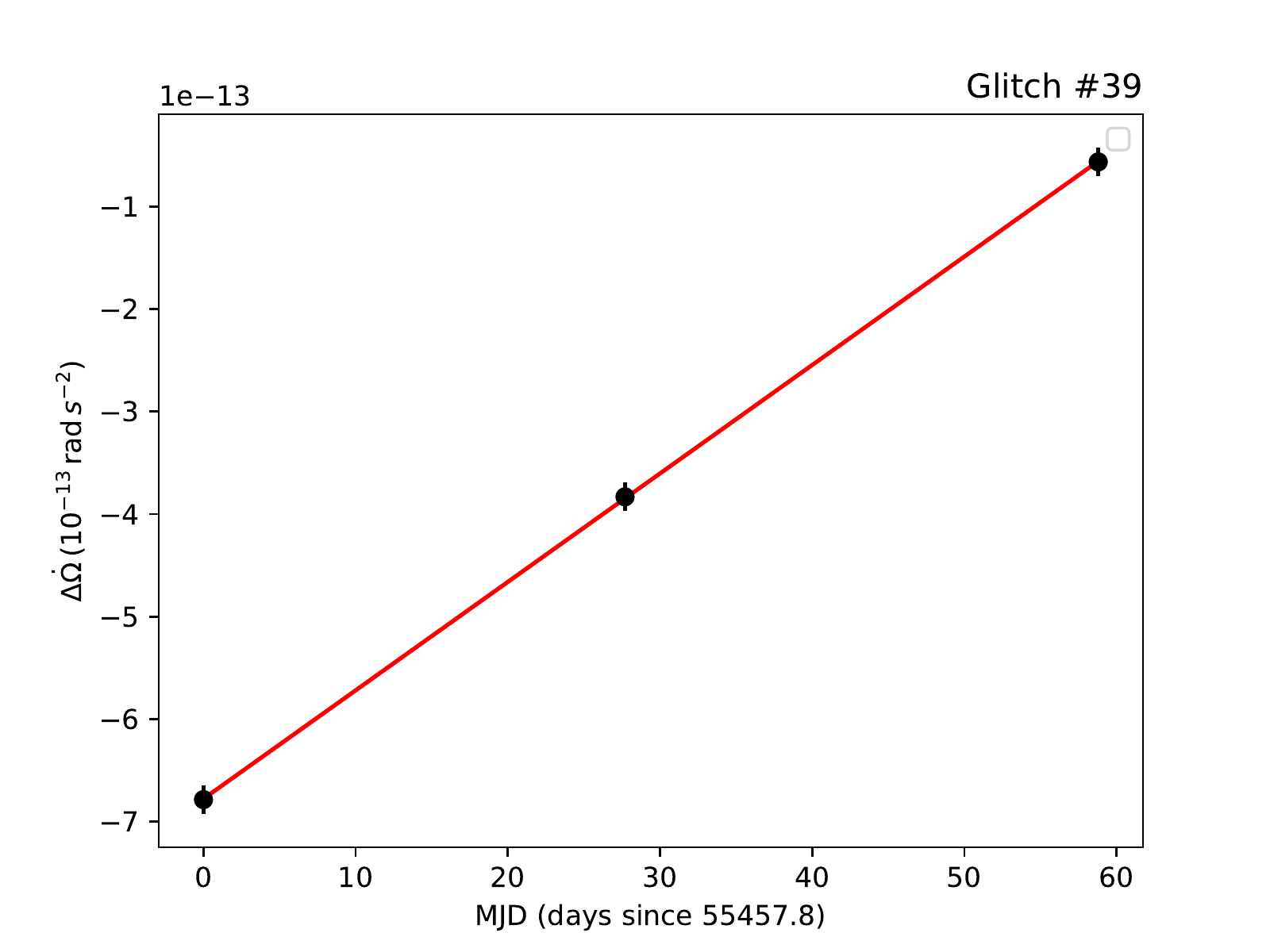}}\

\caption{Model fits to the post-glitch spin-down rate after the 29th, 33th, 34nd, 35th, 36th, 37th, 38th, and 39th glitches of PSR J0537$-$6910.}
\label{some example}
\end{figure*}

\begin{figure*}
\centering

\subfloat[Glitch 40]{\includegraphics[width = 3in,height=5.0cm]{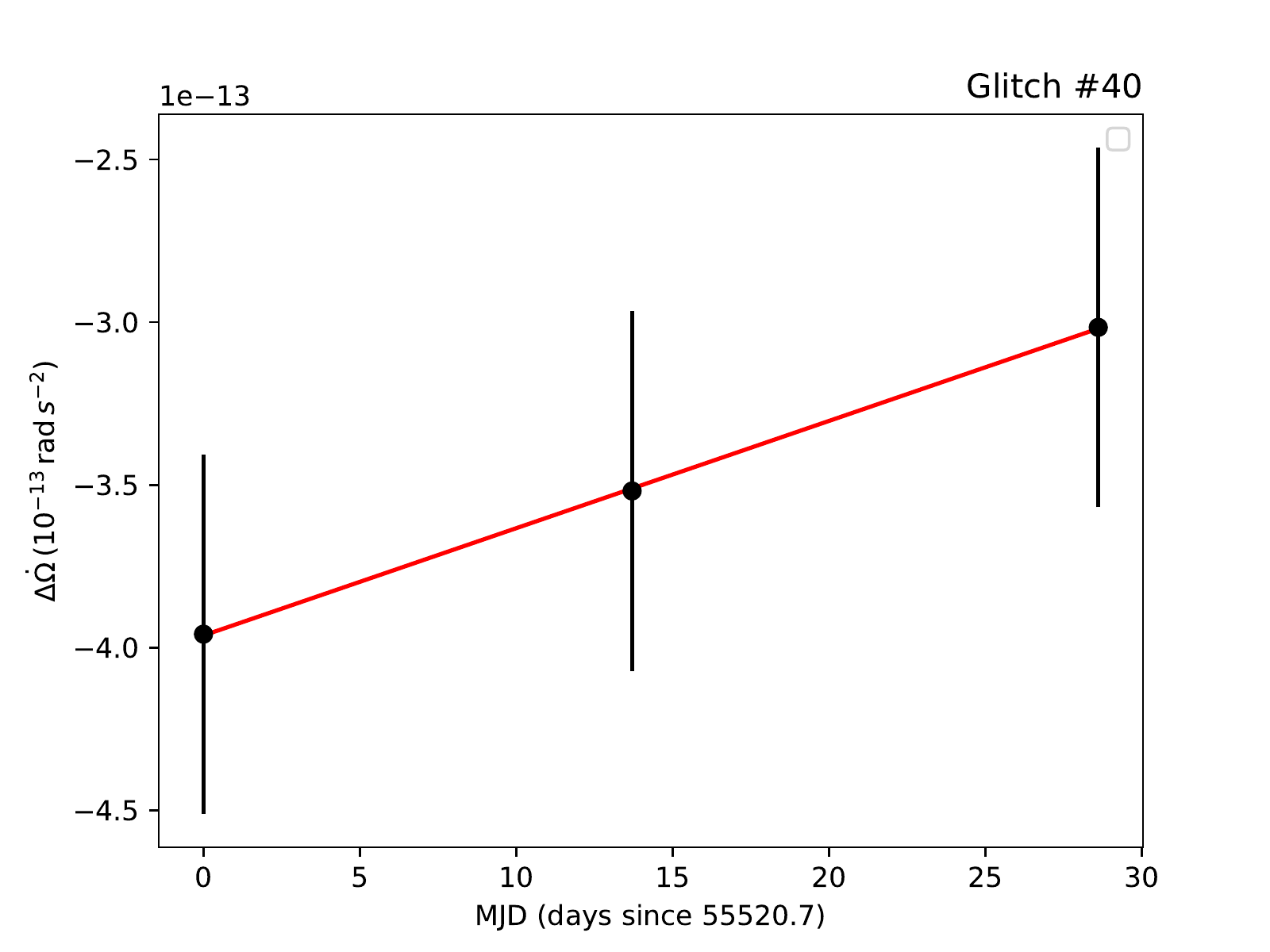}}
\subfloat[Glitch 43]{\includegraphics[width = 3in,height=5.0cm]{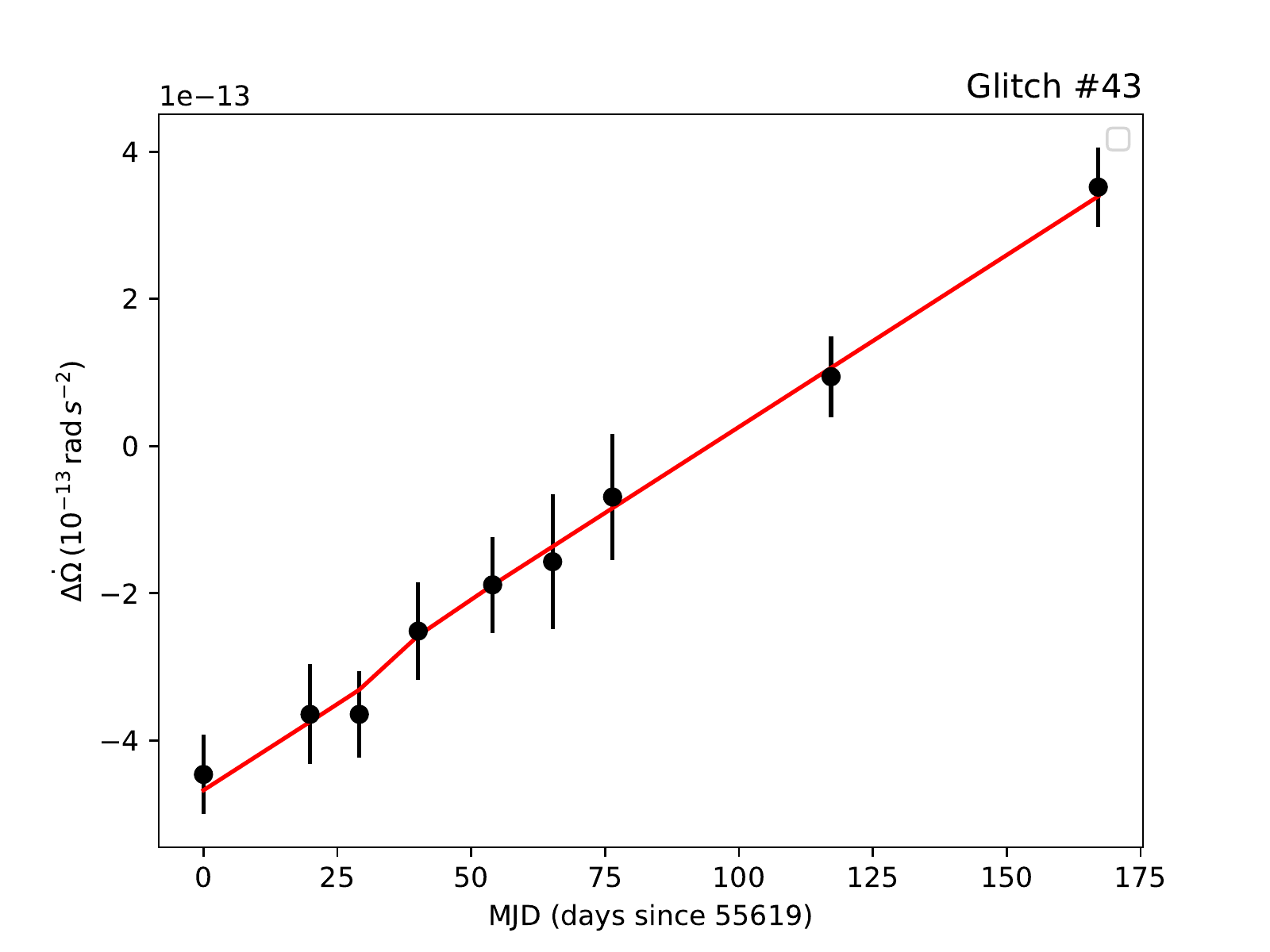}}\
\subfloat[Glitch 45]{\includegraphics[width = 3in,height=5.0cm]{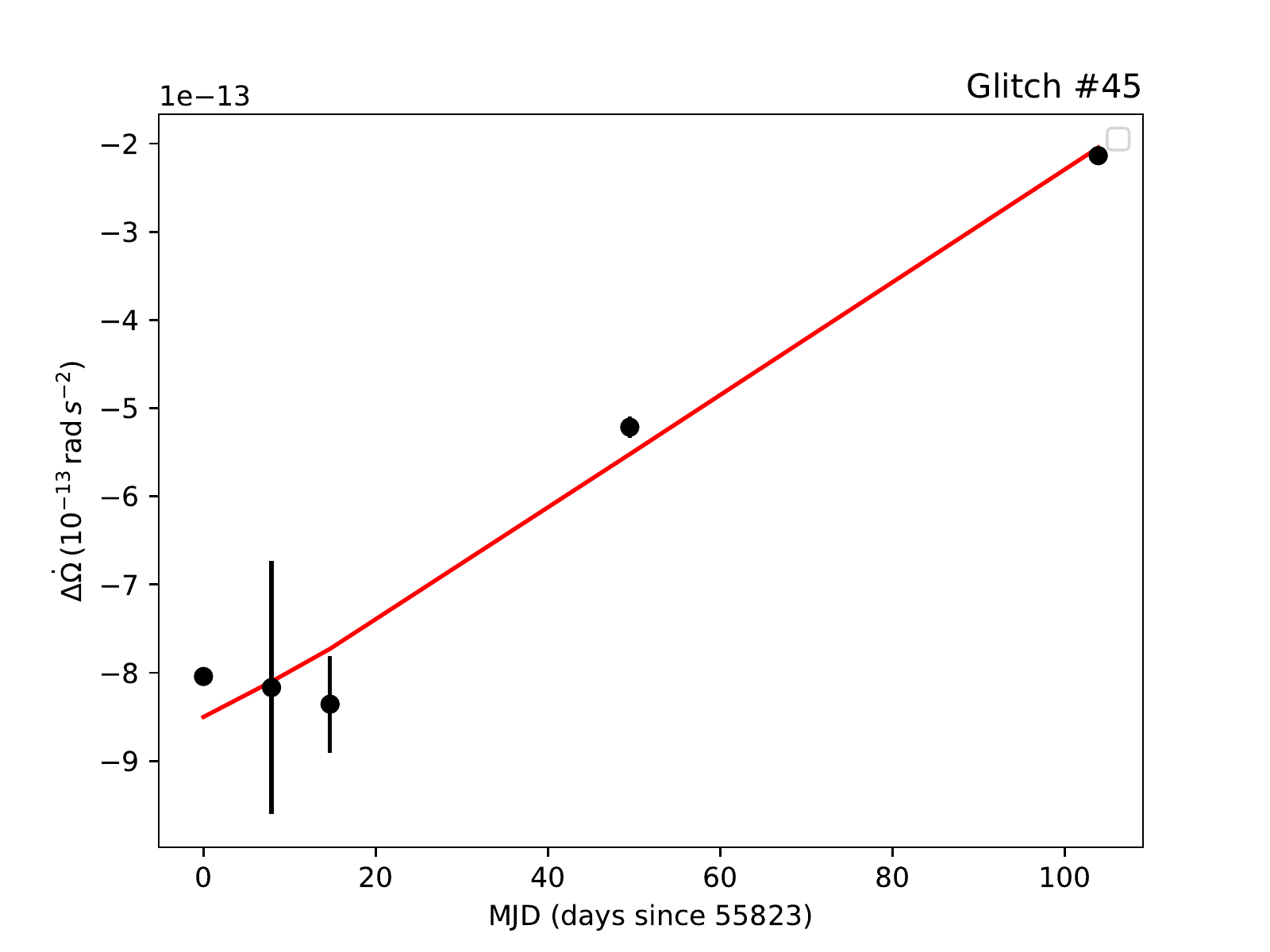}}

\caption{Model fits to the post-glitch spin-down rate after the 40th, 43th, and 45th glitches of PSR J0537$-$6910.}
\label{some example}
\end{figure*}


\bsp	
\label{lastpage}
\end{document}